 \documentclass[final,5p,times,twocolumn]{elsarticle}

 \usepackage{graphics}
 \usepackage{graphicx}
\usepackage{amssymb}
\usepackage{amsmath}

\journal{Journal of Molecular Spectroscopy}

\begin{document}

\begin{frontmatter}



\title{Rotational spectroscopy as a tool to investigate interactions between vibrational 
polyads in symmetric top molecules: low-lying states $\varv_8 \le 2$ of methyl cyanide, CH$_3$CN}


\author[Koeln]{Holger S.P.~M\"uller\corref{cor}}
\ead{hspm@ph1.uni-koeln.de}
\cortext[cor]{Corresponding author.}
\author[JPL]{Linda R. Brown}
\author[JPL]{Brian J. Drouin}
\author[JPL]{John C. Pearson}
\author[Paris]{Isabelle Kleiner}
\author[PNNL]{Robert L. Sams}
\author[JPL]{Keeyoon Sung}
\author[Koeln]{Matthias H. Ordu}
\author[Koeln]{Frank Lewen}

\address[Koeln]{I.~Physikalisches Institut, Universit{\"a}t zu K{\"o}ln, 
  Z{\"u}lpicher Str. 77, 50937 K{\"o}ln, Germany}
\address[JPL]{Jet Propulsion Laboratory, California Institute of Technology, 
  Pasadena, CA 91109-8099, USA}
\address[Paris]{Laboratoire Interuniversitaire des Syst{\`e}mes Atmosph{\'e}riques (LISA), 
  UMR CNRS 7583, Universit{\'e}s Paris Est Cr{\'e}teil et Paris Diderot, Institut Pierre 
  Simon Laplace, 61 Avenue du G{\'e}n{\'e}ral de Gaulle, 94010 Cr{\'e}teil Cedex, France}
\address[PNNL]{Pacific Northwest National Laboratory, P.O. Box 999, Mail Stop K8-88, 
  Richland, WA 99352, USA}

%%%%%%%%%%%%%%%%%%%%%%%%%%%%%%%%%%%%%%%%%%%%%%%%%%%%%%%%%%%%%%%%%%%%%%%%%%%%%%%%%%%%%
%%%%%%%%%%%%%%%%%%%%%%%%%%%%%%%%%%%%%%%%%%%%%%%%%%%%%%%%%%%%%%%%%%%%%%%%%%%%%%%%%%%%%
%%%%%%%%%%%%%%%%%%%%%%%%%%%%%%%%%%%%%%%%%%%%%%%%%%%%%%%%%%%%%%%%%%%%%%%%%%%%%%%%%%%%%
\begin{abstract}

Rotational and rovibrational spectra of methyl cyanide were recorded to analyze 
interactions in low-lying vibrational states and to construct line lists for 
radio astronomical observations in space as well as for infrared spectroscopic 
investigations of planetary atmospheres. The rotational spectra cover large portions 
of the 36$-$1627~GHz region. In the infrared (IR), a spectrum was recorded for 
this study in the region of $2\nu_8$ around 717~cm$^{-1}$ with assignments covering 
684$-$765~cm$^{-1}$. Additional spectra in the $\nu _8$ region were used to validate 
the analysis.

Information on the $K$ level structure of CH$_3$CN is almost exclusively obtained 
from IR spectra, as are basics of the $J$ level structure. The large amount and 
the high accuracy of the rotational data improves knowledge of the $J$ level 
structure considerably. Moreover, since these data extend to much higher $J$ and 
$K$ quantum numbers, they allowed us to investigate for the first time in depth 
local interactions between these states which occur at high $K$ values.
In particular, we have detected several interactions between $\varv_8 = 1$ and 2. 
Notably, there is a strong $\Delta \varv_8 = \pm 1$, $\Delta K = 0$, 
$\Delta l = \pm3$ Fermi resonance between $\varv_8 = 1^{-1}$ and $\varv_8 = 2^{+2}$ 
at $K = 14$. Pronounced effects in the spectrum are also caused by resonant 
$\Delta \varv_8 = \pm 1$, $\Delta K = \mp 2$, $\Delta l = \pm 1$ interactions between 
$\varv_8 = 1$ and 2 at $K = 13$, $l = -1$/$K = 11$, $l = 0$ and at $K = 15$, 
$l = +1$/$K = 13$, $l = +2$. An equivalent resonant interaction occurs 
between $K = 14$ of the ground vibrational state and $K = 12$, $l = +1$ of 
$\varv_8 = 1$ for which we present the first detailed account. A preliminary 
account was given in an earlier study on the ground vibrational state. 
Similar resonances were found for CH$_3$CCH and, more recently, for CH$_3$NC, 
warranting comparison of the results. From data pertaining to $\varv_8 = 2$, 
we also investigated rotational interactions with $\varv_4 = 1$ as well as 
$\Delta \varv_8 = \pm 1$, $\Delta K = 0$, $\Delta l = \pm3$ Fermi interactions 
between $\varv_8 = 2$ and 3.

We have derived N$_2$- and self-broadening coefficients for the $\nu_8$, 
$2\nu_8 - \nu_8$, and $2\nu_8$ bands from previously determined $\nu_4$ values. 
Subsequently, we determined transition moments and intensities for the three IR bands.

\end{abstract}

\begin{keyword}  %%% up to 6 !!

rotational spectroscopy \sep 
infrared spectroscopy \sep
vibration-rotation interaction \sep
methyl cyanide \sep
interstellar molecule


\end{keyword}

\end{frontmatter}

% \linenumbers

%%%%%%%%%%%%%%%%%%%%%%%%%%%%%%%%%%%%%%%%%%%%%%%%%%%%%%%%%%%%%%%%%%%%%%%%%%%%%%%%%%%%%
%%%%%  main text  %%%%%%%%%%%%%%%%%%%%%%%%%%%%%%%%%%%%%%%%%%%%%%%%%%%%%%%%%%%%%%%%%%%
%%%%%%%%%%%%%%%%%%%%%%%%%%%%%%%%%%%%%%%%%%%%%%%%%%%%%%%%%%%%%%%%%%%%%%%%%%%%%%%%%%%%%

%%%%%%%%%%%%%%%%%%%%%%%%%%%%%%%%%%%%%%%%%%%%%%%%%%%%%%%%%%%%%%%%%%%%%%%%%%%%%%%%%%%%%
%%%%%  Introduction  %%%%%%%%%%%%%%%%%%%%%%%%%%%%%%%%%%%%%%%%%%%%%%%%%%%%%%%%%%%%%%%%
%%%%%%%%%%%%%%%%%%%%%%%%%%%%%%%%%%%%%%%%%%%%%%%%%%%%%%%%%%%%%%%%%%%%%%%%%%%%%%%%%%%%%

\section{Introduction}
\label{introduction}

Among the more than 180 different molecules detected in the interstellar medium (ISM) 
or in circumstellar envelopes (CSEs) of late type stars, there are 56 with a CN moiety, 
see, e.g., the Molecules in Space page\footnote{https://cdms.astro.uni-koeln.de/classic/molecules} 
of the Cologne Database for Molecular Spectroscopy, CDMS~\cite{CDMS_1,CDMS_2} or the 
Bibliography of Astromolecules\footnote{http://www.astrochymist.org/astrochymist\_mole.html} 
of the Astrochymist web-page\footnote{http://www.astrochymist.org/}. Almost two thirds 
of these can be viewed as cyanides in a wider sense. The alkyl cyanides 
\textit{n}-propyl cyanide, \textit{n}-C$_3$H$_7$CN, also known as 1-cyanopropane or 
butyronitrile, and \textit{iso}-propyl cyanide, \textit{i}-C$_3$H$_7$CN, also known 
as 2-cyanopropane or \textit{iso}-butyronitrile, both recently detected in the very 
massive and luminous galactic center source Sagittarius B2(N) (Sgr~B2(N) for short) 
\cite{n-PrCN_EtFo-det,i-PrCN_det} are the largest complex organic molecules detected 
in the ISM thus far, and, after HC$_{11}$N, the second largest cyanides. Moreover, 
\textit{iso}-propyl cyanide was the first branched alkyl compound detected in space 
\cite{i-PrCN_det}. Ethyl cyanide, C$_2$H$_5$CN, the next smaller homolog, was detected 
much earlier in the Orion Molecular Cloud as well as Sgr~B2 \cite{EtCN-det}. 
Its $^{13}$C isotopologs were detected in Orion~IRc2 \cite{13C-EtCN-det_Orion-KL} and 
Sgr~B2(N) \cite{13C-EtCN-det_Sgr_B2}. In addition, rotational transitions pertaining 
to the two lowest-lying vibrational states were detected in G327.3$-$0.6 
\cite{vib_EtCN_2000} and Sgr~B2(N) \cite{vib_EtCN_2004}. Transitions of higher 
excited states were detected in molecular clouds of Orion \cite{high-vib_EtCN_2013} 
and Sgr~B2(N) \cite{SgrB2_survey_2013}.

Methyl cyanide, CH$_3$CN, also known as acetonitrile or cyanomethane, is the smallest 
alkyl cyanide. It is a trace constituent in Earth's atmosphere which was studied, e.g., 
by high-resolution spectroscopy using microwave limb sounding \cite{MeCN_MLS_Earth-atmo_2001} 
and balloone-borne IR spectroscopy \cite{MeCN_IR_Earth-atmo_2005}. It is produced mainly 
by biomass burning \cite{biomass_burning_MLS_2004,biomass_burning_2011}. 
Acetonitrile is also a useful solvent in the chemical industry.

Methyl cyanide is a very important molecule in space. First identified in Sgr~A and B 
more than 40 years ago, it was among the first molecules to be observed by radio astronomy 
\cite{MeCN-det}. It was also seen in comets, such as Kohoutek \cite{MeCN-Kohoutek_1974}, 
in the atmosphere of Saturn's moon Titan \cite{MeCN-Titan_1993}, 
in dark clouds, such as TMC-1 \cite{MeCN-TMC-1_1983}, around the low-mass protostar 
IRAS~16293$-$2422 \cite{MeCN_IRAS_16293_2003}, in the CSE of the famous carbon-rich 
late-type star CW~Leonis, also known as IRC~+10216 \cite{MeCN-CSE_2008}, and in external 
galaxies, such as NGC~253 \cite{MeCN_MeCCH_extragal}. Several isotopologs of CH$_3$CN were 
detected as well in the ISM, including the rare CH$_2$DCN \cite{CH2DCN-det} and possibly even 
$^{13}$CH$_3 ^{13}$CN \cite{SgrB2_survey_2013}. Transitions of methyl cyanide 
within its lowest excited vibrational state $\varv_8 = 1$ were detected more than 
30 years ago \cite{MeCN_HC3N_vib_1983}. More recently, brief accounts on 
observations of transitions belonging to $\varv_8 = 2$ 
\cite{SgrB2_survey_2013,MeCN_v8=2_2012}, $\varv_4 = 1$ \cite{SgrB2_survey_2013} 
as well as on transitions of vibrationally excited methyl cyanide with $^{13}$C 
\cite{SgrB2_survey_2013} have been published.

The identifications were based on extensive laboratory investigations which began almost 
70 years ago \cite{MeCN_rot_1947}. The most recent account on the ground state rotational 
spectrum of the main isotopolog CH$_3$CN was given by some of the present authors along with 
analyses for several minor isotopologs \cite{MeCN_rot_2009}; unlabeled atoms refer to $^1$H, 
$^{12}$C, and $^{14}$N. The CH$_3$CN line list was based mainly on an accurate and 
extensive study \cite{MeCN_rot_2006} which employed earlier data for frequencies below 
74~GHz \cite{MeCN_1-0,MeCN-12-13b_2-1,MeCN-Lille_1977}. Information on the purely 
$K$-dependent terms cannot be determined by rotational spectroscopy for a strongly prolate 
symmetric top, such as methyl cyanide, unless from the very weak $\Delta K = 3$ transitions 
that gain intensity from centrifugal distortion effects on the the dipole moment. 
Thus the purely $K$-dependent terms $A$, $D_K$, and an estimate of $H_K$ were determined 
from $\Delta K = 3$ ground state energy differences \cite{MeCN_DeltaK=3_1993}.

The lowest excited vibrational state of CH$_3$CN is the doubly degenerate CCN bending mode 
$\varv_8 = 1$ at a vibrational energy of 365.0~cm$^{-1}$ \cite{MeCN_nu8_1992}. It was studied first 
in 1961 by rotational spectroscopy between 92 and 222~GHz ($J'' = 4$ to 11) \cite{MeCN-v8=1_1961}. 
Soon thereafter, more accurate data with partly resolved $^{14}$N hyperfine structure between 18 and 
148~GHz ($J'' = 0$ to 7) were reported \cite{MeCN-v8=1_1969,Bauer_thesis_1970}. Later, data were 
obtained near 370~GHz ($J'' = 19$) \cite{MeCN-vib_le_v4_J=19_1988} and, employing a laser-sideband 
spectrometer, in selected regions between 589 and 1395~GHz ($J'' = 31$ to 75) \cite{MeCN-v8=1_2_1991}. 
Direct $l$-type transitions between the $k \times l = +1$ components were also observed with 
$21 \leq J \leq 45$ \cite{MeCN-v8=1-l-type_1968,MeCN-v8=1-l-type_etc_1991}.

To our knowledge, there was only one prior high-resolution IR study of the $\nu _8$ band 
\cite{MeCN_nu8_1992}, although there were earlier low- to moderate-resolution studies of 
$\nu _8$ or indirect determinations of the vibrational energy from higher vibrational 
bands. A resonant interaction with $\Delta k = -2$ and $\Delta l = +1$ between $\varv = 0$, 
$K = 14$ and $\varv_8 = 1^{+1}$, $K = 12$ at $J = 43$ was reported \cite{MeCN_rot_2004}, 
but could only be analyzed approximately because only two slightly perturbed transitions 
in the ground vibrational state ($J'' = 38$ and 39) were observed.

The parallel $2\nu _8$ band at 716.7~cm$^{-1}$ \cite{MeCN_2nu8_1993} was treated as an 
isolated state in that study even though a rather strong Fermi interaction of $\varv_8 = 2^0$ 
with $\varv_4 = 1$, the non-degenerate C$-$C stretching mode at 920.3~cm$^{-1}$ ($W _{488} 
\approx 30$~cm$^{-1}$) was proposed from a force field analysis \cite{FF_Duncan_1978}. 
This interaction has not been considered in high resolution analyses of $\varv _8 = 2$ thus far. 
Transitions to $\varv _8 = 2^{\pm2}$ became observable only through $q_{22}$ interaction with 
$\varv _8 = 2^0$ \cite{MeCN_2nu8_1993}. A Fermi interaction between $\varv_7 = 1^{+1}$, 
and $\varv_8 = 2^{-2}$ at $K = 12$ and 13 was proposed to explain deviations in the high-$K$ 
part of the $\nu _7$ band \cite{MeCN_nu7_etc_Laser-Stark_1984}. 
This interaction was confirmed in a later study mainly concerned with $\nu _4$, $\nu _7$ and 
$3 \nu _8 ^1$ bands between 850 and 1150~cm$^{-1}$ \cite{MeCN_nu4_nu7_3nu8_1993}; 
in addition, a $\Delta k = -1$ and $\Delta l = +2$ interaction between $\varv_8 = 2^{-2}$ 
and $\varv_4 = 1$ at $K = 6$ and 5, respectively was found for the first time along with 
several other perturbations among the higher lying states.

The rotational spectrum of CH$_3$CN in its $\varv_8 = 2$ state has also been studied between 
18 and 112~GHz ($J'' = 0$ to 5) \cite{Bauer_thesis_1970,MeCN-v8=2_1969,MeCN-v8=2_1971}, 
near 370~GHz ($J'' = 19$) \cite{MeCN-vib_le_v4_J=19_1988} and, employing a laser-sideband 
spectrometer, in selected regions between 787 and 1273~GHz ($J'' = 20$ to 68) 
\cite{MeCN-v8=1_2_1991}.

Despite the numerous studies of methyl cyanide in various spectral regions, there has been no 
detailed attempt to study its spectroscopy globally. If there were effects, which were caused by 
vibrational states outside of the polyad under investigation, they were only considered through 
the effective parameters, see, e.g., Refs.~\cite{MeCN_rot_2004,MeCN_nu4_nu7_3nu8_1993}.

In the present work, we have recorded extensive rotational and rovibrational spectra in order 
to study vibration-rotation interactions not only within vibrational states or within vibrational 
polyads, but also between vibrational polyads. In comparison to earlier data, we access 
higher $J$ and $K$ quantum numbers, have a more complete coverage of the data and 
typically higher accuracies. The data permitted the first detailed analysis of a resonant 
interaction between $\varv = 0$ and $\varv_8 = 1$ and the first detections and analyses of 
interactions between $\varv_8 = 1$ and 2 as well as between $\varv_8 = 2$ and 3. 
Preliminary results were described earlier \cite{MeCN-OSU_2007}. In this article, we present 
an expanded analysis of these vibrational states of methyl cyanide with $\varv_8$ up to 2 and 
provide preliminary analyses of interactions with vibrational states pertaining to the next polyad. 
We calculate rotational and rovibrational spectra from our resulting analysis to support methyl 
cyanide observations in diverse environments in space as well as in Earth's atmosphere and those 
of other planets or related objects.

Interactions between vibrational polyads of symmetric top molecules, such as methyl cyanide, 
are facilitated by the comparatively small vibrational energy of 365~cm$^{-1}$ for $\varv_8 = 1$ 
combined with the large Coriolis coupling between the two $l$ components which causes levels 
with $\Delta K = 0$, but different $l$ to repel each other increasingly at higher $K$ 
such that the lower energy $l$ component can interact with the ground vibrational state 
at specific $K$ whereas the higher energy $l$ component in turn can interact with the 
different $l$ components of $\varv_8 = 2$ also at specific $K$. Vibrational polyads are groups 
of one or more vibrational states separated by approximately equal amounts of vibrational energy, 
see e.g. Ref.~\cite{polyad-based_review} for a recent review.

Interactions in symmetric top molecules between vibrational polyads were studied first in 
the case of propyne, CH$_3$CCH, also known as methylacetylene, which is isoelectronic to 
methyl cyanide. A $\Delta \varv = \pm 1$, $\Delta K = 0$, $\Delta l = \pm3$ Fermi interaction 
at $K = 12$ between $\varv _{10} = 1^{-1}$ and $\varv _{10} = 2^{+2}$ as well as a Coriolis 
interaction with $\Delta K \mp 1$, $\Delta l = \pm 2$ at $K = 10$ and 9, respectively, 
between $\varv _{10} = 1^{-1}$ and $\varv _{9} = 1^{+1}$ were first revealed mainly by 
rotational spectroscopy \cite{MeCCH_nu10+Dyade_2002} and later improved by infrared as well 
as additional rotational spectroscopy \cite{MeCCH_nu10+Dyade_2004}. The latter study also 
revealed a rotational (Coriolis-type) resonance between $\varv _{10} = 2$ and 
$\varv _5 = 1$, which is the lowest vibration of the next higher tetrad. 
Interactions between the dyad and the tetrad were studied in greater detail later 
\cite{MeCCH_Dyade+Triade_2009}. More recently, such Fermi and $\alpha$-resonances 
were analyzed between $\varv_8 = 1$ and 2 in methyl isocyanide \cite{CH3NC_v8=1_2_2011}. 
The vibrational energy of $\varv_8 = 1$ is lower, only 267.3~cm$^{-1}$, such that the 
Fermi interaction occurs already at $K = 9$ and 10, while they occur at $K = 12$ and 14 
for the corresponding resonances in methylacetylene and methyl cyanide.

We are not aware of resonant interactions between the ground and an excited vibrational 
state in the case of a symmetric top molecule other than the case of CH$_3$CN. 
However, a Coriolis-type interaction between the ground vibrational state of HNCO 
and its lowest $\varv_5 = 1$ state was proposed and its strength estimated 
\cite{HNCO_05_estimate_1980}. An analysis was presented for the analogous case of 
the isoelectronic HN$_3$ molecule \cite{HN3_0-5-6_analysis_1987}. A remarkable case is 
vinyl cyanide, C$_2$H$_3$CN, also known as acrylonitrile, for which the ground vibrational 
state was connected to its lowest $\varv_{11} = 1$ state through $\Delta K = 5$ and 6 
interactions without observation of high-resolution infrared transitions 
\cite{VyCN_0-11_2009}. Recently, $\varv_{11} = 1$ was connected to $\varv_{15} = 1$ 
and $\varv_{11} = 2$ as well as the latter two to each other \cite{VyCN_le22_2012}. 
The density of vibrational states increases with energy, and therefore chances to 
connect states at lower quantum numbers and through interactions with smaller 
$\Delta K$ increase. However, observation may be difficult in practice because 
transitions in higher vibrational states are weaker than corresponding ones in 
lower vibrational states. In addition, the number of at least similarly strong 
rotational transitions per frequency unit increases also with vibrational state.

%%%%%%%%%%%%%%%%%%%%%%%%%%%%%%%%%%%%%%%%%%%%%%%%%%%%%%%%%%%%%%%%%%%%%%%%%%%%%%%%%%%%%
%%%%%  Experimental details  %%%%%%%%%%%%%%%%%%%%%%%%%%%%%%%%%%%%%%%%%%%%%%%%%%%%%%%%
%%%%%%%%%%%%%%%%%%%%%%%%%%%%%%%%%%%%%%%%%%%%%%%%%%%%%%%%%%%%%%%%%%%%%%%%%%%%%%%%%%%%%

\section{Experimental details}
\label{exptl_details}

%%%%%%%%%%%%%%%%%%%%%%%%%%%%%%%%%%%%%%%%%%%%%%%%%%%%%%%%%%%%%%%%%%%%%%%%%%%%%%%%%%%%%
\subsection{Rotational spectra at the Jet Propulsion Laboratory}
%%%%%%%%%%%%%%%%%%%%%%%%%%%%%%%%%%%%%%%%%%%%%%%%%%%%%%%%%%%%%%%%%%%%%%%%%%%%%%%%%%%%%

Most of the CH$_3$CN rotational data were extracted from broad frequency scans taken 
with the JPL cascaded multiplier spectrometer \cite{JPL_multiplier_spectrometer_2005}. 
Generally, a multiplier chain source is passed through a $1-2$ meter pathlength flow cell 
and is detected by a silicon bolometer cooled to near 1.7~K. The cell is filled with a 
steady flow of reagent grade acetonitrile at room temperature, and the pressure and 
modulation are optimized to enable good S/N ratios with narrow lineshapes. 
The S/N ratio was optimized for a higher-$K$ transition (e.g. $K = 12$) 
because of the very strong ground state transitions of the main isotopolog with lower $K$, 
which may exhibit saturated line profiles. 
This procedure enables better dynamic range for the extraction of line positions for rare 
isotopologs and highly excited vibrational satellites. The frequency ranges covered were 
440$-$540, 619$-$631, 638$-$648, 770$-$855, 875$-$930, 967$-$1050, 1083$-$1093, 
1100$-$1159, 1168$-$1198, 1576$-$1591, and 1614$-$1627~GHz, as in our previous analysis 
\cite{MeCN_rot_2009}. In addition, we obtained new data at 530$-$595~GHz and data with 
improved S/N in the 780$-$835~GHz region. Most of these multiplier sources were previously 
described \cite{JPL_multiplier_spectrometer_2005}. However, the multiplier chain with 
frequency range coverage between 967$-$1050~GHz was not described in that work. 
This chain consists of two cascaded triplers after the amplified W-band stage, the peak 
output power is near 100~$\mu$W. The efficiency of frequency multipliers is usually strongly 
frequency-dependent. In addition, recording conditions and sensitivities of detectors can 
have strong influences on the quality of the spectra. Particularly good S/N ratios were 
reached around 600, 800, 900 and at 1100$-$1200~GHz. The S/N ratios changed considerably 
within each scan and were usually lower towards the edges. Uncertainties of 50~kHz were 
assigned to isolated lines with good to moderate S/N ratios, smaller uncertainties, down 
to 10~kHz, in cases of very good S/N and very symmetric lineshapes, larger uncertainties 
if the S/N was poorer or if the lineshape was rather asymmetric, e.g. because of the 
proximity to another line.

%%%%%%%%%%%%%%%%%%%%%%%%%%%%%%%%%%%%%%%%%%%%%%%%%%%%%%%%%%%%%%%%%%%%%%%%%%%%%%%%%%%%%
%%%%%%%%%%%%%%%%%%%%%%%%%%%%%%%%%%%%%%%%%%%%%%%%%%%%%%%%%%%%%%%%%%%%%%%%%%%%%%%%%%%%%
\subsection{Rotational spectra at the Universit{\"at} zu K{\"o}ln}
%%%%%%%%%%%%%%%%%%%%%%%%%%%%%%%%%%%%%%%%%%%%%%%%%%%%%%%%%%%%%%%%%%%%%%%%%%%%%%%%%%%%%
%%%%%%%%%%%%%%%%%%%%%%%%%%%%%%%%%%%%%%%%%%%%%%%%%%%%%%%%%%%%%%%%%%%%%%%%%%%%%%%%%%%%%

All measurements at the Universit{\"a}t zu K{\"o}ln were recorded at room temperature 
in static mode employing different Pyrex glass cells having an inner diameter of 
100~mm with pressures in the range of 0.1$-$0.5~Pa at lower frequencies, increased 
to around 1.0~Pa at higher frequencies. The window material was Teflon at lower 
frequencies, whereas high-density polyethylene was used at higher frequencies.
Frequency modulation was used throughout with demodulation at $2f$, causing an isolated 
line to appear close to a second derivative of a Gaussian.

The $J = 2 - 1$ and $3 - 2$ transitions of several vibrational states around 37 
and 55~GHz, respectively, were recorded with an Agilent E8257D microwave synthesizer 
as source and a home-built Schottky diode detector. A 7~m long double pass absorption 
cell was used for these measurements. All other measurements were carried out in 
shorter (around 3~m) single pass cells. 
The $J = 4 - 3$ transitions of $\varv _8 = 2$ around 74.0~GHz were recorded using 
a backward-wave oscillator (BWO) based 4~mm synthesizer AM-MSP~1 (Analytik \& 
Me{\ss}technik GmbH, Chemnitz, Germany) as source and a Schottky-diode as detector. 
A small number of methyl cyanide rotational transitions were recorded around 850~GHz 
and around 880~GHz with the Cologne Terahertz Spectrometer \cite{CTS1994} using a BWO 
as source and a liquid helium cooled InSb hot-electron bolometer (QMC) as detector. 
Further, extensive measurements were carried out in parts of the 1330$-$1501~GHz 
region employing a frequency multiplier (Virginia Diode, Inc.) driven by an Agilent 
E8257D microwave synthesizer as source and the InSb bolometer as detector. 
This spectrometer was used previously for investigations of CH$_3$SH and 
$^{13}$CH$_3$OH \cite{MeSH_rot_2012,13CH3OH_rot_2014}.

%%%%%%%%%%%%%%%%%%%%%%%%%%%%%%%%%%%%%%%%%%%%%%%%%%%%%%%%%%%%%%%%%%%%%%%%%%%%%%%%%%%%%
%%%%%  Figure 1  %%%%%%%%%%%%%%%%%%%%%%%%%%%%%%%%%%%%%%%%%%%%%%%%%%%%%%%%%%%%%%%%%%%%
%%%%%%%%%%%%%%%%%%%%%%%%%%%%%%%%%%%%%%%%%%%%%%%%%%%%%%%%%%%%%%%%%%%%%%%%%%%%%%%%%%%%%

 \begin{figure}
 \begin{center}
  \includegraphics[angle=-90,width=8.8cm]{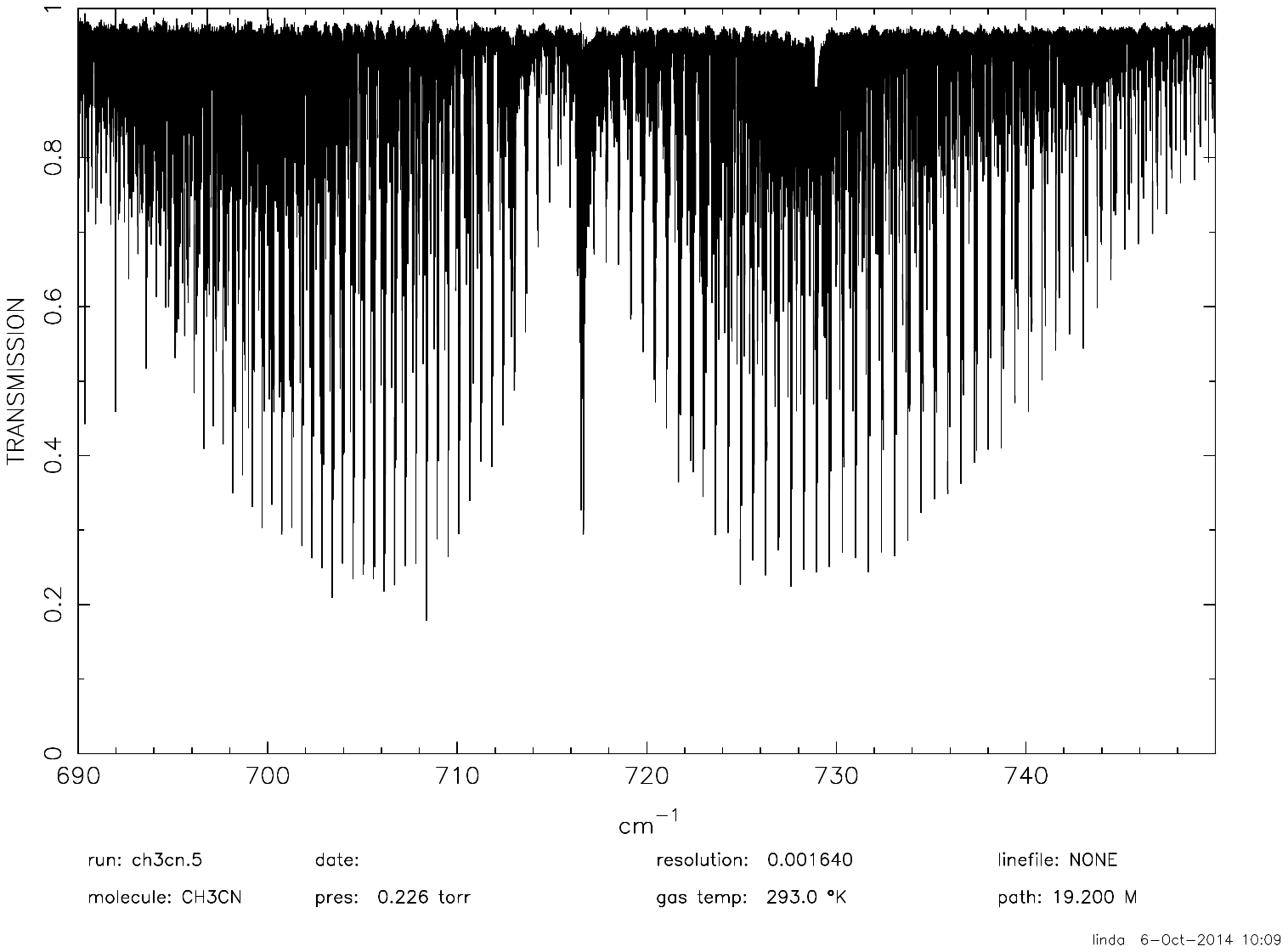}
 \end{center}
  \caption{The 2$\nu_8$ spectrum of CH$_3$CN recorded at 0.0016~cm$^{-1}$ resolution 
           using a Bruker 120~HR FTS at PNNL; the optical path is 19.2~m, and the gas 
           pressure is 0.226~Torr at 293~K. The sample included $\sim$0.4\,\% CO$_2$ 
           and $\sim$3\,\% OCS for frequency calibration.}
  \label{2nu8_overview}
 \end{figure}

%%%%%%%%%%%%%%%%%%%%%%%%%%%%%%%%%%%%%%%%%%%%%%%%%%%%%%%%%%%%%%%%%%%%%%%%%%%%%%%%%%%%%
%%%%%  Figure 2  %%%%%%%%%%%%%%%%%%%%%%%%%%%%%%%%%%%%%%%%%%%%%%%%%%%%%%%%%%%%%%%%%%%%
%%%%%%%%%%%%%%%%%%%%%%%%%%%%%%%%%%%%%%%%%%%%%%%%%%%%%%%%%%%%%%%%%%%%%%%%%%%%%%%%%%%%%

 \begin{figure}
 \begin{center}
  \includegraphics[angle=-90,width=8.8cm]{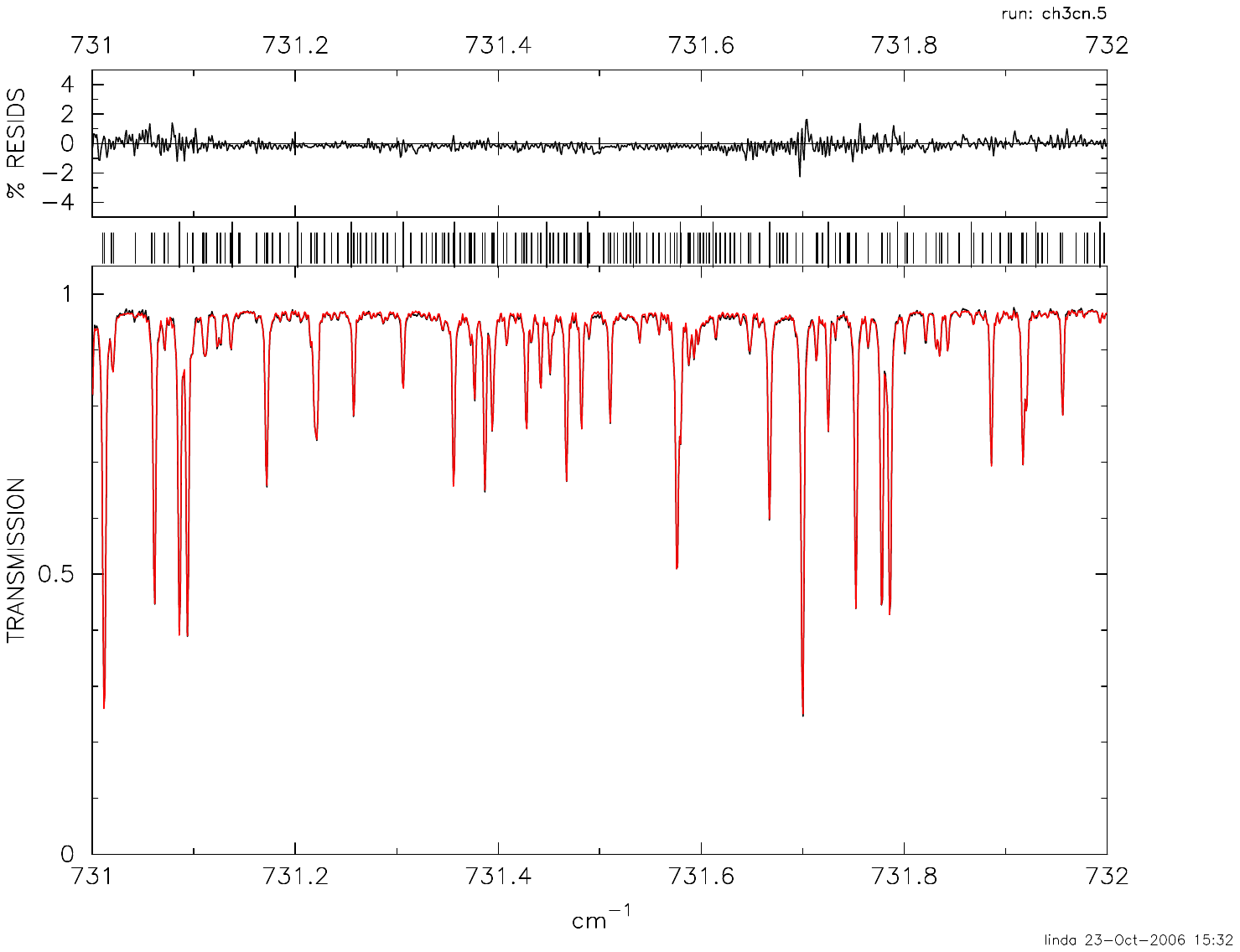}
 \end{center}
  \caption{Comparison of CH$_3$CN lines at 731.5~cm$^{-1}$ retrieved by least squares 
           curve-fitting. The lower panel has the observed spectrum in black overlaid 
           by the synthetic spectrum in red. The upper panel shows the residual 
           difference (calc.-obs.) in percent. The optical path is 19.2~m, and the 
           gas pressure is 0.226~Torr at 293~K.The vertical lines between the two 
           panels mark the measured features (every 10$^{\rm th}$ line is taller).}
  \label{2nu8_good_positions}
 \end{figure}

%%%%%%%%%%%%%%%%%%%%%%%%%%%%%%%%%%%%%%%%%%%%%%%%%%%%%%%%%%%%%%%%%%%%%%%%%%%%%%%%%%%%%
%%%%%%%%%%%%%%%%%%%%%%%%%%%%%%%%%%%%%%%%%%%%%%%%%%%%%%%%%%%%%%%%%%%%%%%%%%%%%%%%%%%%%
%%%%%%%%%%%%%%%%%%%%%%%%%%%%%%%%%%%%%%%%%%%%%%%%%%%%%%%%%%%%%%%%%%%%%%%%%%%%%%%%%%%%%

\subsection{Infrared spectrum}
\label{exptl_IR}

One infrared spectrum of CH$_3$CN was recorded between 600 and 989~cm$^{-1}$ at 
0.0016~cm$^{-1}$ resolution using an MCT detector with a Bruker 120~HR Fourier transform 
spectrometer at Pacific Northwest National Laboratory (PNNL); the 2$\nu_8$ region is 
shown in \textbf{Fig.~\ref{2nu8_overview}}. For this, a multi-pass absorption cell, 
set to an optical path length of 19.20~m, was filled to 0.226~Torr (30.1~Pa) of CH$_3$CN 
at 293.0~K. Small amounts of OCS and CO$_2$ were added to the sample for frequency 
calibration; comparison of 63 well-isolated lines of the $\nu_3$ fundamental of OCS 
at 860~cm$^{-1}$ \cite{OCS_ref_1992} produced a calibration factor 1.000000868~(21) 
with an rms of 0.0000182~cm$^{-1}$. Nearly 12000 line positions and relative intensities 
between 690 and 765~cm$^{-1}$ were retrieved using non-linear least-squares curve-fitting 
\cite{curve-fitting_IR-positions_1983}. However, the corresponding precisions in the 
2$\nu_8$ region were an order of magnitude worse than the OCS calibration because the 
observed spectrum was so congested. \textbf{Fig.~\ref{2nu8_good_positions}} shows 
the quality of the curve fitting for a small section of $2\nu _8$.

Two other room-temperature spectra spanned the $\nu_8$ region (250 to 400~cm$^{-1}$) 
at 294~K using pressures of 0.78~Torr (103.9~Pa) with a 19.2~m path and 1.01~Torr 
(134.5~Pa) with a 3.24~m path. These were calibrated using residual water inside the FTS 
\cite{HITRAN_2012} to obtain a calibration factor of 1.000000207 (495) with an rms of 
only 0.000140~cm$^{-1}$. The spectra were very congested and no attempt was made 
to retrieve postions and intensities by curve-fitting. Previously reported CH$_3$CN 
line centers \cite{MeCN_nu8_1992} had better precisions in this second region, and so 
these new $\nu_8$ data were used mainly to search for hot bands and to confirm assignments 
in the 2$\nu_8$ region.

%%%%%%%%%%%%%%%%%%%%%%%%%%%%%%%%%%%%%%%%%%%%%%%%%%%%%%%%%%%%%%%%%%%%%%%%%%%%%%%%%%%%%
%%%%%  Observed spectrum  %%%%%%%%%%%%%%%%%%%%%%%%%%%%%%%%%%%%%%%%%%%%%%%%%%%%%%%%%%%
%%%%%%%%%%%%%%%%%%%%%%%%%%%%%%%%%%%%%%%%%%%%%%%%%%%%%%%%%%%%%%%%%%%%%%%%%%%%%%%%%%%%%

\section{Results}
\label{results}

%%%%%%%%%%%%%%%%%%%%%%%%%%%%%%%%%%%%%%%%%%%%%%%%%%%%%%%%%%%%%%%%%%%%%%%%%%%%%%%%%%%%%
\subsection{Overview of CH$_3$CN spectroscopy}
%%%%%%%%%%%%%%%%%%%%%%%%%%%%%%%%%%%%%%%%%%%%%%%%%%%%%%%%%%%%%%%%%%%%%%%%%%%%%%%%%%%%%
\label{intro-spec}

%%%%%%%%%%%%%%%%%%%%%%%%%%%%%%%%%%%%%%%%%%%%%%%%%%%%%%%%%%%%%%%%%%%%%%%%%%%%%%%%%%%%%
%%%%%  Figure 3  %%%%%%%%%%%%%%%%%%%%%%%%%%%%%%%%%%%%%%%%%%%%%%%%%%%%%%%%%%%%%%%%%%%%
%%%%%%%%%%%%%%%%%%%%%%%%%%%%%%%%%%%%%%%%%%%%%%%%%%%%%%%%%%%%%%%%%%%%%%%%%%%%%%%%%%%%%

 \begin{figure}
 \begin{center}
  \includegraphics[width=5.0cm]{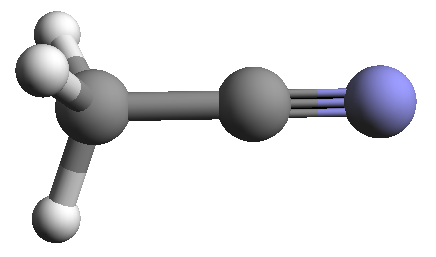}
 \end{center}
  \caption{Model of the methyl cyanide molecule. The C atoms are shown in gray, 
           the H atoms in light gray, and the N atom is shown in blue. The $a$-axis is 
           along the CCN atoms and is also the symmetry axis.}
  \label{CH3CN_molecule}
 \end{figure}

%%%%%%%%%%%%%%%%%%%%%%%%%%%%%%%%%%%%%%%%%%%%%%%%%%%%%%%%%%%%%%%%%%%%%%%%%%%%%%%%%%%%%
%%%%%   Table 1  %%%%%%%%%%%%%%%%%%%%%%%%%%%%%%%%%%%%%%%%%%%%%%%%%%%%%%%%%%%%%%%%%%%%
%%%%%%%%%%%%%%%%%%%%%%%%%%%%%%%%%%%%%%%%%%%%%%%%%%%%%%%%%%%%%%%%%%%%%%%%%%%%%%%%%%%%%

\begin{table}
\begin{center}
\caption{Energies (cm$^{-1}$) and symmetries Sym of low-lying vibrational states of methyl cyanide.}
\label{vib_energies}
% \smallskip
{\footnotesize
\begin{tabular}[t]{lcr@{}ll}
\hline 
State                     & Sym. & \multicolumn{2}{c}{Energy} & Reference                          \\
\hline
$\varv = 0$               & $A$  &          0&.0              & per definitionem                   \\
$\varv_8 = 1$             & $E$  &        365&.024            & this work                          \\
$\varv_8 = 2^0$           & $A$  &        716&.750            & this work                          \\
$\varv_8 = 2^2$           & $E$  &        739&.148            & this work                          \\
$\varv_4 = 1$             & $A$  &        920&.290            & Ref.~\cite{MeCN_nu4_nu7_3nu8_1993} \\
$\varv_7 = 1$             & $E$  &       1041&.855            & Ref.~\cite{MeCN_nu4_nu7_3nu8_1993} \\
$\varv_8 = 3^1$           & $E$  &       1077&.786            & Ref.~\cite{MeCN_nu4_nu7_3nu8_1993} \\
$\varv_8 = 3^3$           & $A$  & $\sim$1122&.15             & Ref.~\cite{MeCN_nu4_nu7_3nu8_1993} \\
$\varv_4 = \varv_8 = 1$   & $E$  &       1290&.0              & Ref.~\cite{pentade_1994}           \\
$\varv_3 = 1$             & $A$  &       1385&.2              & Ref.~\cite{pentade_1994}           \\
$\varv_7 = \varv_8 = 1^0$ & $A$  &       1401&.7              & Ref.~\cite{pentade_1994}           \\
$\varv_7 = \varv_8 = 1^2$ & $E$  &       1408&.9              & Ref.~\cite{pentade_1994}           \\
$\varv_8 = 4^0$           & $A$  & $\sim$1426&.               & this work                          \\
$\varv_8 = 4^2$           & $E$  &       1447&.9              & Ref.~\cite{pentade_1994}           \\
$\varv_6 = 1$             & $E$  &       1449&.7              & Ref.~\cite{pentade_1994}           \\
$\varv_8 = 4^4$           & $E$  & $\sim$1514&.               & this work                          \\
\hline \hline
\end{tabular}\\[2pt]
}
\end{center}
\end{table}

%%%%%%%%%%%%%%%%%%%%%%%%%%%%%%%%%%%%%%%%%%%%%%%%%%%%%%%%%%%%%%%%%%%%%%%%%%%%%%%%%%%%%
%%%%%%%%%%%%%%%%%%%%%%%%%%%%%%%%%%%%%%%%%%%%%%%%%%%%%%%%%%%%%%%%%%%%%%%%%%%%%%%%%%%%%

As shown in \textbf{Fig.~\ref{CH3CN_molecule}}, the light H atoms are the only ones 
not on the symmetry axis of CH$_3$CN. Hence, methyl cyanide is a strongly prolate 
molecule ($A \gg B$). Rotational transitions with $\Delta K = 0$ are strong 
because of the large dipole moment of 3.92197~(13)~D \cite{MeCN-dipole}. 
$\Delta K = 3$ transitions only gain intensity through centrifugal 
distortion effects and are usually very weak. In the absence of perturbations, 
the purely axial parameters $A$ (or $A - B$), $D_K$, etc. cannot be determined by 
rotational spectroscopy. Moreover, even rovibrational spectroscopy yields, strictly 
speaking, the differences $\Delta A$ (or $\Delta (A - B)$), $\Delta D_K$, etc. 
from single state analyses, but not the ground state parameters. In the case of 
CH$_3$CN, analyses of three IR bands involving two doubly degenerate vibrational 
modes, $\nu_8$, $\nu_7 + \nu_8$, and $\nu_7 + \nu_8 - \nu_8$, permitted the axial 
parameters to be determined \cite{MeCN_DeltaK=3_1993}.

Methyl cyanide has four totally symmetric and four doubly degenerate vibrational modes. 
The three lowest energy fundamentals are the doubly degenerate CCN bending mode $\nu_8$ 
at 365.024~cm$^{-1}$ \cite{MeCN_nu8_1992}, the totally symmetric CC stretching mode 
$\nu_4$ at 920.290~cm$^{-1}$ \cite{MeCN_nu4_nu7_3nu8_1993}, and the doubly degenerate 
CH$_3$ rocking mode $\nu_7$ at 1041.855~cm$^{-1}$ \cite{MeCN_nu4_nu7_3nu8_1993}. 
They are comparatively weak with integrated room temperature cross sections in the 
range 2 to $5 \times 10^{-19}$~cm/molecule \cite{MeCN-int_2005,MeCN-int_1995}. 
Interestingly, the $2\nu_8$ overtone band is slightly stronger than $\nu_8$ or 
$\nu_4$ \cite{MeCN-int_2005,MeCN-int_1995}. In the case of $\nu_4$, line-by-line 
intensity measurements have also been published \cite{MeCN_nu4_int_2008}.
A list of low-lying vibrational states, up to $\sim$1500~cm$^{-1}$, is given in 
\textbf{Table~\ref{vib_energies}}.

%%%%%%%%%%%%%%%%%%%%%%%%%%%%%%%%%%%%%%%%%%%%%%%%%%%%%%%%%%%%%%%%%%%%%%%%%%%%%%%%%%%%%
%%%%%  Figure 4  %%%%%%%%%%%%%%%%%%%%%%%%%%%%%%%%%%%%%%%%%%%%%%%%%%%%%%%%%%%%%%%%%%%%
%%%%%%%%%%%%%%%%%%%%%%%%%%%%%%%%%%%%%%%%%%%%%%%%%%%%%%%%%%%%%%%%%%%%%%%%%%%%%%%%%%%%%

 \begin{figure}
 \begin{center}
  \includegraphics[width=8.8cm]{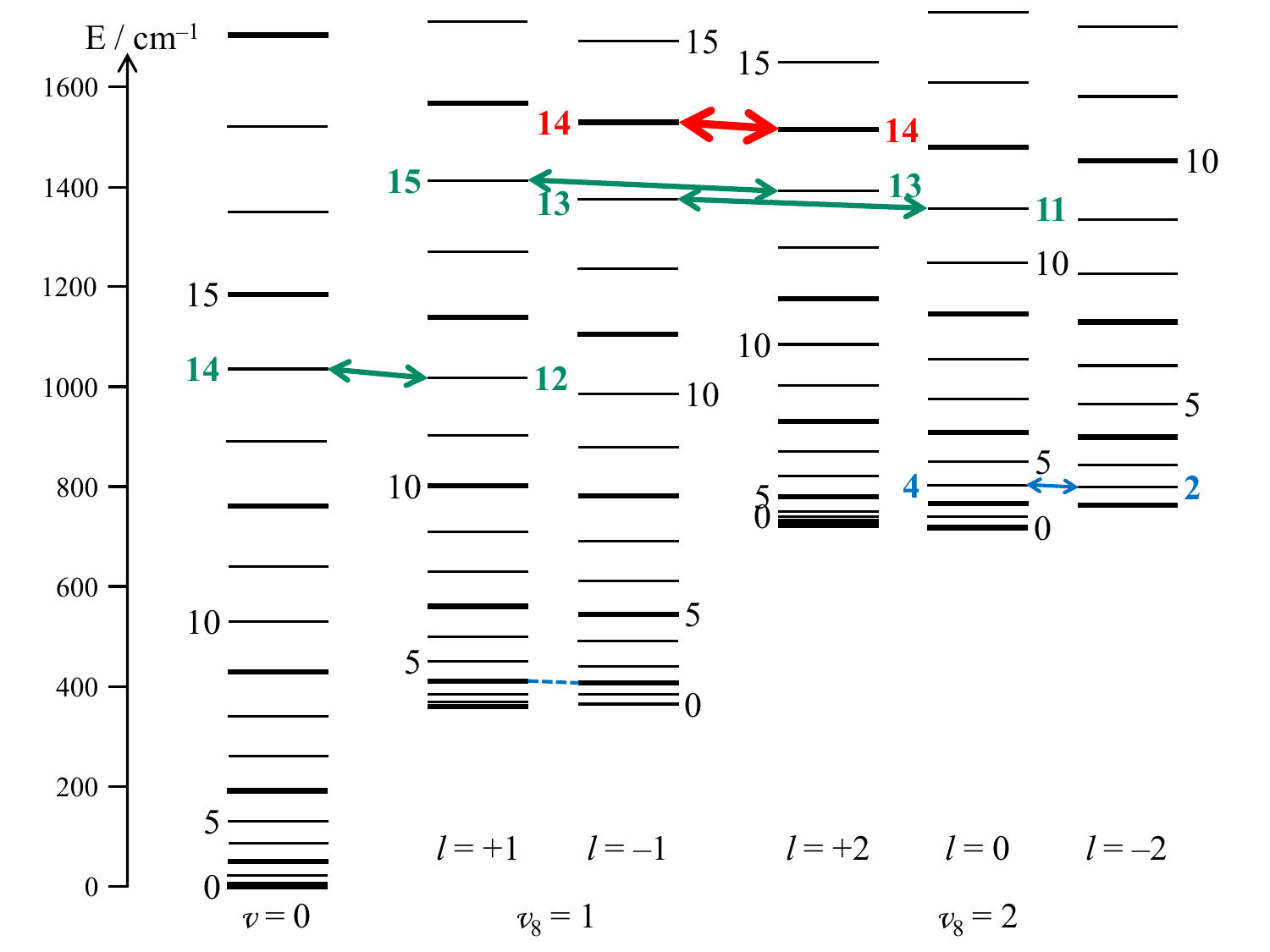}
 \end{center}
  \caption{$K$ level structure of methyl cyanide displaying the $J = K$ energies for 
           $\varv _8 = 0$, 1, and 2 with their $l$ substates. Levels with $A$ symmetry 
           are drawn with thick lines. Coriolis coupling within $\varv _8 = 1$ causes 
           near-degeneracies of levels having $\Delta K = \Delta l = 2$ and hence strong 
           $q_{22}$ interaction shown exemplarily for one pair of low $K$ by a thin, 
           dashed (blue) line. The same applies to $\varv _8 = 2^{\pm2}$, where a 
           $q_{22}$ resonance occurs between $K = 4$ and $K = 2$ of the $l = 0$ and 
           $-$2 components, respectively (thin blue left right arrow). 
           The spacing between $K$ levels is fairly large compared to the vibrational 
           spacing causing a $\Delta K = 0$, $\Delta l = 3$ Fermi interaction at 
           $K = 14$ between $\varv _8 = 1^{-1}$ and $\varv _8 = 2^{+2}$, displayed 
           by a thick (red) left right arrow. Also shown are $\Delta K = \mp2$, 
           $\Delta l = \pm1$ interactions between $\varv _8 = 0$ and 1 as well as 
           between $\varv _8 = 1$ and 2 (medium thick green left right arrow).}
  \label{v8_le_2_K-level}
 \end{figure}

%%%%%%%%%%%%%%%%%%%%%%%%%%%%%%%%%%%%%%%%%%%%%%%%%%%%%%%%%%%%%%%%%%%%%%%%%%%%%%%%%%%%%
%%%%%%%%%%%%%%%%%%%%%%%%%%%%%%%%%%%%%%%%%%%%%%%%%%%%%%%%%%%%%%%%%%%%%%%%%%%%%%%%%%%%%
%%%%%  Figure 5  %%%%%%%%%%%%%%%%%%%%%%%%%%%%%%%%%%%%%%%%%%%%%%%%%%%%%%%%%%%%%%%%%%%%
%%%%%%%%%%%%%%%%%%%%%%%%%%%%%%%%%%%%%%%%%%%%%%%%%%%%%%%%%%%%%%%%%%%%%%%%%%%%%%%%%%%%%

 \begin{figure}
 \begin{center}
  \includegraphics[width=4.4cm]{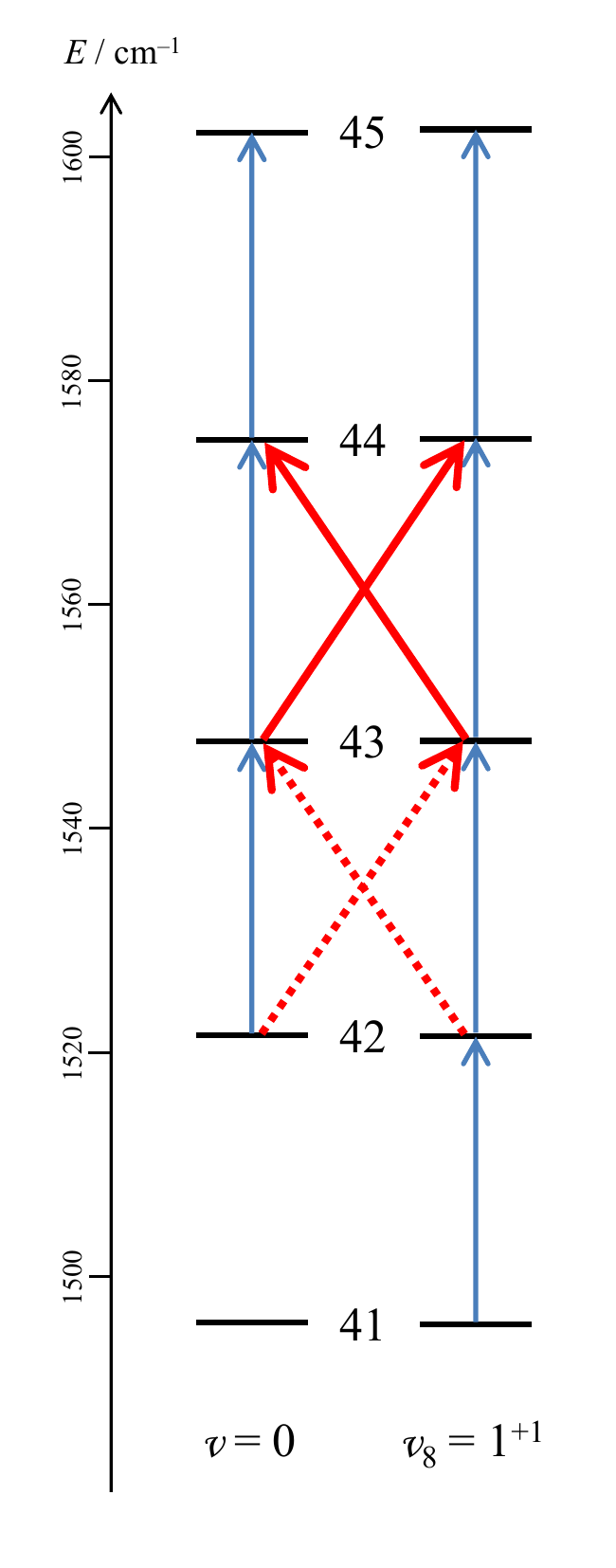}
 \end{center}
  \caption{Detail of the energy level structure of methyl cyanide around the perturbation 
           between the ground state and $\varv _8 = 1$. Transitions observable in our 
           spectral recordings are represented by arrows; those with solid lines 
           indicate transitions in the final line list, including two transitions between 
           the vibrational states. Two further transitions between $\varv = 0$ and 
           $\varv _8 = 1$ with significant intensities, but overlapped with stronger 
           transitions are displayed by arrows with dotted lines.}
  \label{gs_v8_perturbation}
 \end{figure}

%%%%%%%%%%%%%%%%%%%%%%%%%%%%%%%%%%%%%%%%%%%%%%%%%%%%%%%%%%%%%%%%%%%%%%%%%%%%%%%%%%%%%
%%%%%%%%%%%%%%%%%%%%%%%%%%%%%%%%%%%%%%%%%%%%%%%%%%%%%%%%%%%%%%%%%%%%%%%%%%%%%%%%%%%%%

The fairly large value of $A$, $\sim$5.27~cm$^{-1}$, leads to a rapid increase in rotational 
energy with $K$ which is shown up to $\sim$1700~cm$^{-1}$ in \textbf{Fig.~\ref{v8_le_2_K-level}}. 
The highest $K$ levels observed involve $K = 21$, and the $J = K = 21$ energy is at 
2313~cm$^{-1}$.

The vibrational energy of the lowest, doubly degenerate vibrational mode $\varv _8 = 1$ is 
at 365.024~cm$^{-1}$ \cite{MeCN_nu8_1992}, similar to the $J = K = 8$ energy in $\varv = 0$ 
(\textbf{Fig.~\ref{v8_le_2_K-level}}). The Coriolis interaction between the two $l$ 
components of $\varv _8 = 1$, described by $\zeta \approx 0.877$, is large, close to the 
limiting case $\zeta = 1$. It effectively shifts the levels of the $l = +1$ ladder to 
lower energies and the energies of the $l = -1$ ladder to higher energies. As a result, 
$K = 1$ of $l = +1$ is actually lower than $K = 0$. 
More importantly, levels with $\Delta K = \Delta l = 2$ are near-degenerate, leading to 
pronounced $q_{22}$ interaction within $\varv _8 = 1$. The interaction is highlighted in 
\textbf{Fig.~\ref{v8_le_2_K-level}} for $K = 4$ of $\varv _8 = 1^{+1}$ and $K = 2$ of 
$\varv _8 = 1^{-1}$. As a consequence of the Coriolis interaction within $\varv _8 = 1$, 
the $K$ level spacing in $\varv _8 = 1^{+1}$ is considerably smaller than in $\varv = 0$ 
while it is larger in $\varv _8 = 1^{-1}$. \v{S}ime\v{c}kov{\'a} et al. \cite{MeCN_rot_2004} 
found evidence of a resonant $\Delta \varv_8 = \pm 1$, $\Delta K = \mp 2$, $\Delta l = \pm 1$ 
interaction between $\varv = 0$, $K = 14$ and $\varv_8 = 1^{+1}$, $K = 12$ 
(\textbf{Fig.~\ref{v8_le_2_K-level}}) at $J = 43$. They analyzed the interaction approximately 
because only two slightly perturbed transitions in the ground vibrational state 
($J'' = 38$ and 39) were observed.

Other global interactions within an isolated degenerate vibrational mode of a strongly 
prolate rotor are rare, in contrast to nearly spherical or oblate rotors, see, e.g. 
Ref.~\cite{C3v_reduction_2014} and references therein.

The basic features of the $K$ level structure of $\varv _8 = 2$ are similar to those of 
$\varv _8 = 1$, the main difference being three $l$ components, $l = 0, \pm2$ instead of two. 
Pronounced $q_{22}$ interaction occurs also within $\varv _8 = 2$, in particular between 
$l = 0$ and $l = -2$ at low $K$ with a resonant interaction $K = 4$ of $\varv _8 = 2^{0}$ 
and $K = 2$ of $\varv _8 = 2^{-2}$, as shown in \textbf{Fig.~\ref{v8_le_2_K-level}}. 
In addition, the $K$ level structure in $\varv _8 = 2^{+2}$ is irregular at low $K$ with 
$K = 2$ being lowest in energy and $K = 3$ still slightly below $K = 0$. Such irregular $K$ 
level structure in states such as $\varv _8 \ge 2$ may lead to \textit{local} perturbations 
within the vibrational state with, e.g., $\Delta K = \pm 1$, $\Delta l = \mp 2$ ($q_{21}$),  
$\Delta K = \pm 3$, $\Delta l = 0$ ($\epsilon$), etc. In the case of several different 
such local perturbations, indeterminacies, as investigated in Ref.~\cite{C3v_reduction_2014} 
and references therein, may need to be considered. A global effect may be caused by a 
$\Delta K = \pm 4$, $\Delta l = \pm 4$ interaction ($f_{44}$). However, this effect is 
rather small in a strongly prolate rotor, and $f_{44}$ was barely determined for 
$\varv _{10} = 2$ of propyne \cite{MeCCH_nu10+Dyade_2004}.

%%%%%%%%%%%%%%%%%%%%%%%%%%%%%%%%%%%%%%%%%%%%%%%%%%%%%%%%%%%%%%%%%%%%%%%%%%%%%%%%%%%%%
%%%%%   Table 2  %%%%%%%%%%%%%%%%%%%%%%%%%%%%%%%%%%%%%%%%%%%%%%%%%%%%%%%%%%%%%%%%%%%%
%%%%%%%%%%%%%%%%%%%%%%%%%%%%%%%%%%%%%%%%%%%%%%%%%%%%%%%%%%%%%%%%%%%%%%%%%%%%%%%%%%%%%

\begin{table*}
\begin{center}
\caption{Maximum $J$ and $K$ values and number of rotational and IR lines in vibrational 
         substates of methyl cyanide used in the present / previous study$^a$, with 
         numbers retained in the present fit from previous data in parentheses and 
         rms error for subsets of data.}
\label{statistics}
% \smallskip
{\footnotesize
\begin{tabular}[t]{lcccccc}
\hline 
                         & $\varv = 0$       & $\varv _8 = 1^{+1}$ & $\varv _8 = 1^{-1}$ & $\varv _8 = 2^0$  & $\varv _8 = 2^{+2}$ & $\varv _8 = 2^{-2}$ \\
\hline
$J_{\rm max}$(rot)       & 89/89 (89)        & 88/75 (45)          & 88/75 (8)           & 88/68 (20)        & 88/68 (20)          & 88/68 (20)          \\
$K_{\rm max}$(rot)       & 21/21 (21)        & 19/13 (7)           & 17/11 (7)           & 19/12 (12)        & 13/8 (7)            & 11/8 (6)            \\
no. of rot lines$^{b}$   & 316/225 (196)     & 596/c (58)          & 527/c (37)          & 424/85+5$^d$ (18) & 546/73+6$^d$ (24)   & 299/45+5$^d$ (9)    \\
rms error$^e$            & 0.859/0.911       & 0.807/$-$           & 0.796/$-$           & 0.982/$-$         & 0.915/$-$           & 0.957/$-$           \\
$J_{\rm max}$(IR)        & $-$/$-$ ($-$)$^f$ & 71/71 (71)          & 74/74 (74)          & 66/65             & 53/37               & 66/57               \\
$K_{\rm max}$(IR)        & 7/7 (7)           & 11/11 (11)          & 11/11 (11)          & 12/10             &  2/4                &  7/4                \\
no. of  IR lines$^{b}$   & 5/5 (5)$^f$       & 836/g (836)         & 861/g (861)         & 935/944 (0)       & 40/44$^h$ (0)       & 197/122$^h$ (0)     \\
rms error$^e$            & 0.716/0.643       & 0.826/$-$           & 0.797/$-$           & 0.771/$-$         & 0.849/$-$           & 0.727/$-$           \\
\hline \hline
\end{tabular}\\[2pt]
}
\end{center}
{\footnotesize
$^a$ Previous data from Refs.~\cite{MeCN_rot_2009,MeCN_nu8_1992,MeCN_2nu8_1993} and references therein for $\varv = 0$, $\varv _8 = 1$, and 
     $\varv _8 = 2$, respectively.\\
$^b$ Each blend of lines counted as one line. Hyperfine splitting in the present and previous data was considered in the present fit. 
     Transitions with $K = 0$ of $l$-doubled states associated with $\varv _8 = 1^{-1}$ and $\varv _8 = 2^{+2}$, respectively, in the present data set. 
     Lines, which were weighted out, are not counted. Transitions between vibrational states were counted for higher vibrational state.\\
$^c$ Only total number given as 187.\\
$^d$ Numbers after the plus were given zero weight in Ref.~\cite{MeCN_2nu8_1993}.\\ 
$^e$ No rms error was given in Refs.~\cite{MeCN_nu8_1992,MeCN_2nu8_1993}; we assume parameter uncertainties are based on standard errors, 
     i.e. the rms error is 1.0 by definition. The rms errors for data from Ref.~\cite{MeCN_rot_2009} were calculated from the experimental data 
     available at http://www.astro.uni-koeln.de/site/vorhersagen/daten/CH3CN/CH3CN/gs/.\\
$^f$ No individual $\Delta K = 3$ loops were given in Ref.~\cite{MeCN_DeltaK=3_1993}. We used the five $\Delta K = 3$ splittings from Table~II 
     in that work with the reported uncertainties.\\
$^g$ Only total number given as 1705. Slight difference to our numbers mainly because of blended lines, see subsubsection~\ref{nu8_results}.\\
$^h$ 52 assigned $2\nu _8 - \nu _8$ hot band transitions are mentioned in Ref.~\cite{MeCN_2nu8_1993}; see subsubsection~\ref{hot-band-results}.\\
}
\end{table*}

%%%%%%%%%%%%%%%%%%%%%%%%%%%%%%%%%%%%%%%%%%%%%%%%%%%%%%%%%%%%%%%%%%%%%%%%%%%%%%%%%%%%%
\subsection{$\varv = 0$}
%%%%%%%%%%%%%%%%%%%%%%%%%%%%%%%%%%%%%%%%%%%%%%%%%%%%%%%%%%%%%%%%%%%%%%%%%%%%%%%%%%%%%
\label{ground-state}

Recently, we have revisited the ground state rotational spectrum of several isotopic 
species of methyl cyanide \cite{MeCN_rot_2009}. Besides new data, ground state transition 
frequencies for the main species were mainly taken from Ref.~\cite{MeCN_rot_2006} 
with additional lower frequency data \cite{MeCN_1-0,MeCN-12-13b_2-1,MeCN-Lille_1977}. 
The $J$ and $K$ quantum numbers extended to 89 and 21, respectively, see also 
\textbf{Table~\ref{statistics}}. 

The first objective and the main achievement of the present study with respect to the 
ground vibrational state was the detailed analysis of the resonance between $\varv = 0$ 
and $\varv _8 = 1$ mentioned in the previous subsection. Absorption features pertaining 
to $\varv = 0$, $K = 14$, $J'' = 42$ to 49 and to $\varv_8 = 1^{+1}$, $K = 12$, 
$J'' = 41$ to 45 and 47 to 49 could easily be assigned in our spectral recordings 
because of their intensities. Moreover, the perturbed transitions showed regular 
and mostly small displacements from predictions which did not take the interaction 
into account. The largest shifts were $\sim$24~MHz and $\sim$16~MHz for the $J = 43 - 42$ 
and $44 - 43$ transitions, respectively. After these data had been included in the fit, 
two pairs of cross ladder transitions between the two vibrational states were predicted 
to have sufficient intensities for observation. The weaker pair with $J = 44 - 43$ 
were identified, but the ones with $J = 43 - 42$, predicted to be stronger by about 
a factor of 2.5 than the former ones, were blended with stronger lines. 
\textbf{Fig.~\ref{gs_v8_perturbation}} shows a detail of the energy levels around 
the levels perturbed most by the interaction between $\varv = 0$ and $\varv_8 = 1$ 
as well as the observed transitions.

The measurements in K{\"o}ln between 1.33 and 1.5~THz closed a considerable part of 
the gap between $\sim$1.2 and $\sim$1.6~THz in the previous data set \cite{MeCN_rot_2009}. 
We reanalyzed also the spectra between $\sim$1.1 and 1.2~THz taken at JPL and realized 
that the uncertainties were very similar to those of Ref.~\cite{MeCN_rot_2006}, albeit 
with a slightly better coverage and extending to slightly higher $K$. Therefore, we 
employed the JPL data in our present analysis.

We also attempted to assign rotational transitions with $\Delta K = 3$. Their 
predicted uncertainties were small enough to identify them easily if the distortion 
dipole moment was as large as $10^{-4}$~D. However, even then they could be blended 
with stronger methyl cyanide transitions, of which several are not yet assigned. 
No assignments could be made, leading us to conclude that the distortion dipole 
moment is probably smaller. It is also possible that the calculated uncertainties 
of these transitions are too small.

%%%%%%%%%%%%%%%%%%%%%%%%%%%%%%%%%%%%%%%%%%%%%%%%%%%%%%%%%%%%%%%%%%%%%%%%%%%%%%%%%%%%%
\subsection{$\varv _8 = 1$}
%%%%%%%%%%%%%%%%%%%%%%%%%%%%%%%%%%%%%%%%%%%%%%%%%%%%%%%%%%%%%%%%%%%%%%%%%%%%%%%%%%%%%
\label{v8=1}

%%%%%%%%%%%%%%%%%%%%%%%%%%%%%%%%%%%%%%%%%%%%%%%%%%%%%%%%%%%%%%%%%%%%%%%%%%%%%%%%%%%%%
\subsubsection{Rotational data}
%%%%%%%%%%%%%%%%%%%%%%%%%%%%%%%%%%%%%%%%%%%%%%%%%%%%%%%%%%%%%%%%%%%%%%%%%%%%%%%%%%%%%

The rotational spectrum of CH$_3$CN in its $\varv _8 = 1$ lowest vibrational state 
($E_{\rm vib} = 365.02$~cm$^{-1}$) was reported on several occasions 
\cite{MeCN-v8=1_1969,Bauer_thesis_1970,MeCN-vib_le_v4_J=19_1988,MeCN-v8=1_2_1991}. 
The $J'' = 19$ data \cite{MeCN-vib_le_v4_J=19_1988} had residuals in part much larger 
than the reported 50$-$150~kHz uncertainties already in the initial report, which 
were confirmed here. Therefore, and because of their more limited $K$ range, we 
omitted these data from our line list. The laser side-band measurements 
\cite{MeCN-v8=1_2_1991} were also quite limited in $K$ and in accuracy (around 1~MHz) 
and were superseded by our more accurate data. Part of the lower frequency data 
up to 150~GHz \cite{MeCN-v8=1_1969,Bauer_thesis_1970} was also superseded by our data, 
the remainder was retained in the fit with reported uncertainties. We also used in our 
fit direct $l$-type transitions between the $k \times l = +1$ 
components \cite{MeCN-v8=1-l-type_1968,MeCN-v8=1-l-type_etc_1991}.

The interaction between $\varv = 0$ and $\varv _8 = 1$ was already described in 
subsection~\ref{ground-state}. Interactions of $\varv _8 = 1$ with any higher 
vibrational states have not been reported in the literature prior to the present 
investigation.

%%%%%%%%%%%%%%%%%%%%%%%%%%%%%%%%%%%%%%%%%%%%%%%%%%%%%%%%%%%%%%%%%%%%%%%%%%%%%%%%%%%%%
%%%%%  Figure 6  %%%%%%%%%%%%%%%%%%%%%%%%%%%%%%%%%%%%%%%%%%%%%%%%%%%%%%%%%%%%%%%%%%%%
%%%%%%%%%%%%%%%%%%%%%%%%%%%%%%%%%%%%%%%%%%%%%%%%%%%%%%%%%%%%%%%%%%%%%%%%%%%%%%%%%%%%%

 \begin{figure}
 \begin{center}
  \includegraphics[width=8.8cm]{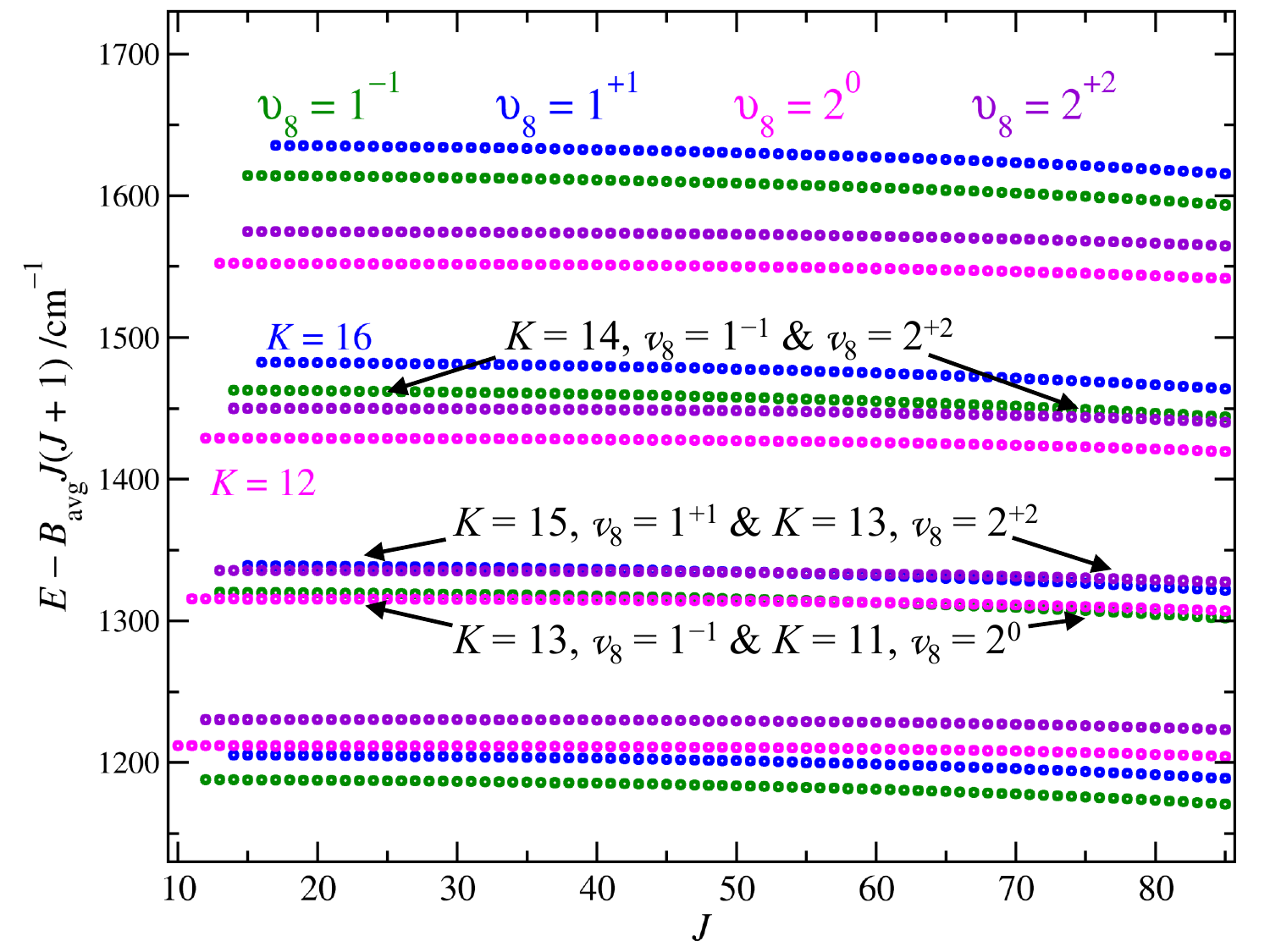}
 \end{center}
  \caption{Reduced energy plot in the region of the perturbations between 
           $\varv _8 = 1$ and 2. A $\Delta \varv_8 = \pm 1$, $\Delta K = 0$, 
           $\Delta l = \pm3$ Fermi resonance occurs near 1450~cm$^{-1}$. 
           In addition, two resonant $\Delta \varv_8 = \pm 1$, $\Delta K = \mp 2$, 
           $\Delta l = \pm 1$ interactions are indicated near 1330~cm$^{-1}$.}
  \label{red_energy_v8_1_2}
 \end{figure}

%%%%%%%%%%%%%%%%%%%%%%%%%%%%%%%%%%%%%%%%%%%%%%%%%%%%%%%%%%%%%%%%%%%%%%%%%%%%%%%%%%%%%
%%%%%%%%%%%%%%%%%%%%%%%%%%%%%%%%%%%%%%%%%%%%%%%%%%%%%%%%%%%%%%%%%%%%%%%%%%%%%%%%%%%%%

Based on the studies on propyne \cite{MeCCH_nu10+Dyade_2002,MeCCH_nu10+Dyade_2004}, we 
expected an interaction between $\varv _8 = 1^{-1}$ and $\varv _8 = 2^{+2}$ at $K$ 
slightly higher than 12 because of the slightly larger vibrational energy difference in 
methyl cyanide compared to propyne, 374.12~cm$^{-1}$ compared to 340.41~cm$^{-1}$. 
Inspection of the energy levels of these CH$_3$CN vibrational states indicated a 
resonant interaction to occur at $K = 14$, see \textbf{Figs.~\ref{v8_le_2_K-level} 
and \ref{red_energy_v8_1_2}}. Using the value of the main Fermi parameter from the 
propyne study \cite{MeCCH_nu10+Dyade_2002,MeCCH_nu10+Dyade_2004} was a reasonable 
initial guess along with published parameters for $\varv _8 = 2$. 
Our final results suggest the interaction to be strongest at $J = 91$ and 92, 
however, the highest assignments presently reach $J = 65 - 64$ for either state 
at these rather high values of $K$. The magnitudes of the perturbations between 
$\varv _8 = 1^{-1}$ and $\varv _8 = 2^{+2}$ showed strong variations with $K$, 
as expected, but also with $J$, most likely because the energy differences between 
the two vibrational ladders also change significantly with $J$. The perturbations 
exceed 1~MHz at $J = 24 - 23$, $K = 10$ and reach 37~MHz at $K = 14$. 
At $J = 65 - 64$, perturbations in the spectrum larger than 1~MHz were 
already seen at $K = 5$, and the shift exceeded 300~MHz at $K = 14$.

\begin{figure}
 \begin{center}
  \includegraphics[width=8.8cm]{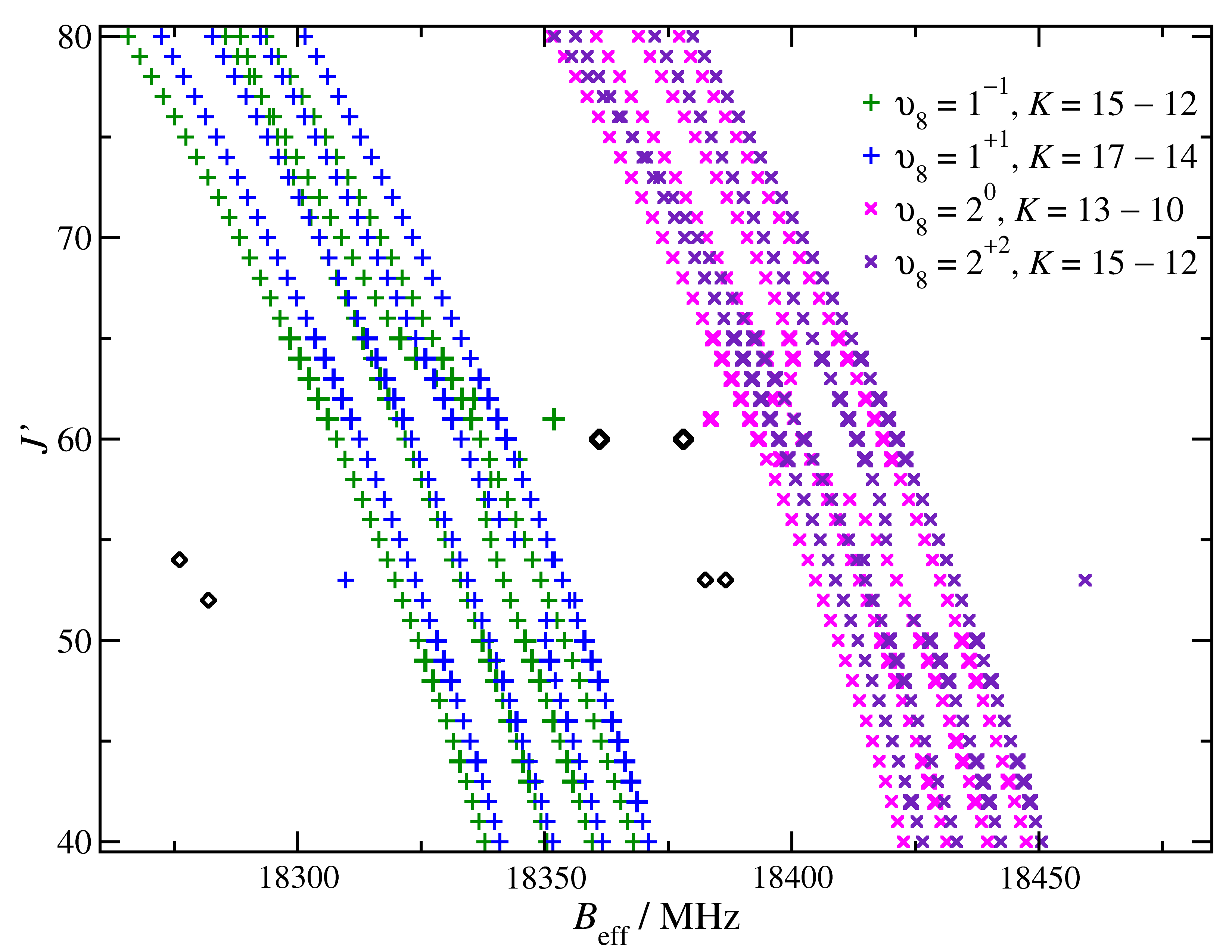}
 \end{center}
  \caption{Fortrat diagram in the region of the resonant $\Delta \varv_8 = \pm 1$, 
           $\Delta K = \mp 2$, $\Delta l = \pm 1$ interactions between 
           $\varv _8 = 1$ and 2. Observed transitions are indicated by larger symbols.}
  \label{Interactions_v8_1_2}
 \end{figure}

The data set up to $J = 28 - 27$ revealed perturbations in $K = 15$ of $\varv _8 = 1^{+1}$ 
which were not accounted for by this $\Delta \varv_8 = \pm 1$, $\Delta K = 0$, 
$\Delta l = \pm3$ Fermi interaction. They were caused by a $\Delta \varv_8 = \pm 1$, 
$\Delta K = \mp 2$, $\Delta l = \pm 1$ interaction with $K = 13$ of $\varv _8 = 2^{+2}$ 
(\textbf{Figs.~\ref{v8_le_2_K-level} and \ref{red_energy_v8_1_2}}), the same type of interaction 
as between $\varv = 0$ and $\varv _8 = 1$. It is shown in \textbf{Fig.~\ref{Interactions_v8_1_2}} 
that the strongest mixing occurs at $J = 52$ and 53. Among the four transitions 
with $J = 53 - 52$, the $\varv _8 = 1$ transition was too close to the lower frequency 
limit of its experimental spectral window, the $\varv _8 = 2$ transition was observed 
with a total perturbation of 1.8~GHz. In addition, both cross ladder transitions were 
observed, but one was too close to a stronger line to determine a reliable and sufficiently 
accurate transition frequency. In addition, the $J = 54 - 53$ transition frequencies 
were determined for both vibrational states.

%%%%%%%%%%%%%%%%%%%%%%%%%%%%%%%%%%%%%%%%%%%%%%%%%%%%%%%%%%%%%%%%%%%%%%%%%%%%%%%%%%%%%
%%%%%  Figure 7  %%%%%%%%%%%%%%%%%%%%%%%%%%%%%%%%%%%%%%%%%%%%%%%%%%%%%%%%%%%%%%%%%%%%
%%%%%%%%%%%%%%%%%%%%%%%%%%%%%%%%%%%%%%%%%%%%%%%%%%%%%%%%%%%%%%%%%%%%%%%%%%%%%%%%%%%%%

%%%%%%%%%%%%%%%%%%%%%%%%%%%%%%%%%%%%%%%%%%%%%%%%%%%%%%%%%%%%%%%%%%%%%%%%%%%%%%%%%%%%%
%%%%%%%%%%%%%%%%%%%%%%%%%%%%%%%%%%%%%%%%%%%%%%%%%%%%%%%%%%%%%%%%%%%%%%%%%%%%%%%%%%%%%

The same type of interaction showed noticeable effects in the spectrum at higher $J$ 
(up to 49) between $K = 13$ of $\varv _8 = 1^{-1}$ and $K = 11$ of $\varv _8 = 2^0$, 
see also \textbf{Figs.~\ref{v8_le_2_K-level} and \ref{red_energy_v8_1_2}}, with the largest 
perturbation at $J = 60$, as shown in \textbf{Fig.~\ref{Interactions_v8_1_2}}. 
The regular rotational transitions were detected for both vibrational states 
in the case of $J = 61 - 60$, both cross-ladder transitions were detected for 
$J = 60 - 59$. The total perturbations amounted to more than 1.5~GHz for the former 
two transitions and more than 3.0~GHz for the latter two. The remaining associated 
transitions were about an order of magnitude weaker and were not identified.

We mention for completeness a $\Delta \varv_8 = \pm 2$, $\Delta K = \mp 2$, 
$\Delta l = \pm 4$ interaction between $K = 15$ of $\varv _8 = 1^{-1}$ and $K = 13$ 
of $\varv _8 = 3^{+3}$ whose strongest effects are currently predicted at $J = 73/74$. 
Should the interaction cause any measureable perturbation in the spectrum, it should 
be very localized because of the high order of the interaction. 
A $\Delta \varv_8 = \pm 2$, $\Delta K = \mp 1$, $\Delta l = \pm 2$ interaction is 
possible between $K = 26$ of $\varv _8 = 1^{-1}$ and $K = 25$ of $\varv _8 = 3^{+1}$, 
but these levels are so high in energy that they are unlikely to be accessed.

The overall coverage of our $\varv _8 = 1$ is indicated in \textbf{Table~\ref{statistics}}. 
Besides the few cross-ladder transitions between $\varv _8 = 1$ and 2, the data are 
limited almost entirely to regular rotational transitions. Transitions between different 
$l$ components may occur because of $\Delta l = \Delta K = 2$ mixing facilitated by $q$. 
Such transitions were predicted to be very weak, and only one transition was 
identified, so we view this identification as tentative. We also searched for 
$\Delta K = 3$ transitions, but, as in the case of the ground vibrational state, 
were unable to identify any with sufficient certainty.

%%%%%%%%%%%%%%%%%%%%%%%%%%%%%%%%%%%%%%%%%%%%%%%%%%%%%%%%%%%%%%%%%%%%%%%%%%%%%%%%%%%%%
\subsubsection{The $\nu_8$ band}
%%%%%%%%%%%%%%%%%%%%%%%%%%%%%%%%%%%%%%%%%%%%%%%%%%%%%%%%%%%%%%%%%%%%%%%%%%%%%%%%%%%%%
\label{nu8_results}

An investigation of $\nu _8$ was reported in Ref.~\cite{MeCN_nu8_1992}, and we use 
the line list from that work as indicated in subsection~\ref{exptl_IR}. Lines with full 
weight were assigned uncertainties of 0.0002~cm$^{-1}$, except for the small number of 
$^rQ_{10}$ lines, for which 0.0004~cm$^{-1}$ were used because of their persistent 
larger residuals in the fits. Lines with weight 0.1 were assigned uncertainties of 
0.0006~cm$^{-1}$. A few lines (around 10 to 20) with weight 0.1 were omitted either 
because of very large residuals or because the line was blended, but was the weaker 
line, and the stronger line was not in the line list. In other rare cases, blended 
lines ($12 \times 2$) were treated as intensity-weighted averages. The effects caused 
by these differences in the previous and present line list on the parameter values, 
uncertainties and on the quality of the fit are probably negligible.

%%%%%%%%%%%%%%%%%%%%%%%%%%%%%%%%%%%%%%%%%%%%%%%%%%%%%%%%%%%%%%%%%%%%%%%%%%%%%%%%%%%%%
\subsection{$\varv _8 = 2$}
%%%%%%%%%%%%%%%%%%%%%%%%%%%%%%%%%%%%%%%%%%%%%%%%%%%%%%%%%%%%%%%%%%%%%%%%%%%%%%%%%%%%%
\label{v8=2}

%%%%%%%%%%%%%%%%%%%%%%%%%%%%%%%%%%%%%%%%%%%%%%%%%%%%%%%%%%%%%%%%%%%%%%%%%%%%%%%%%%%%%
\subsubsection{Rotational data}
%%%%%%%%%%%%%%%%%%%%%%%%%%%%%%%%%%%%%%%%%%%%%%%%%%%%%%%%%%%%%%%%%%%%%%%%%%%%%%%%%%%%%

\begin{figure}
 \begin{center}
  \includegraphics[width=8.8cm]{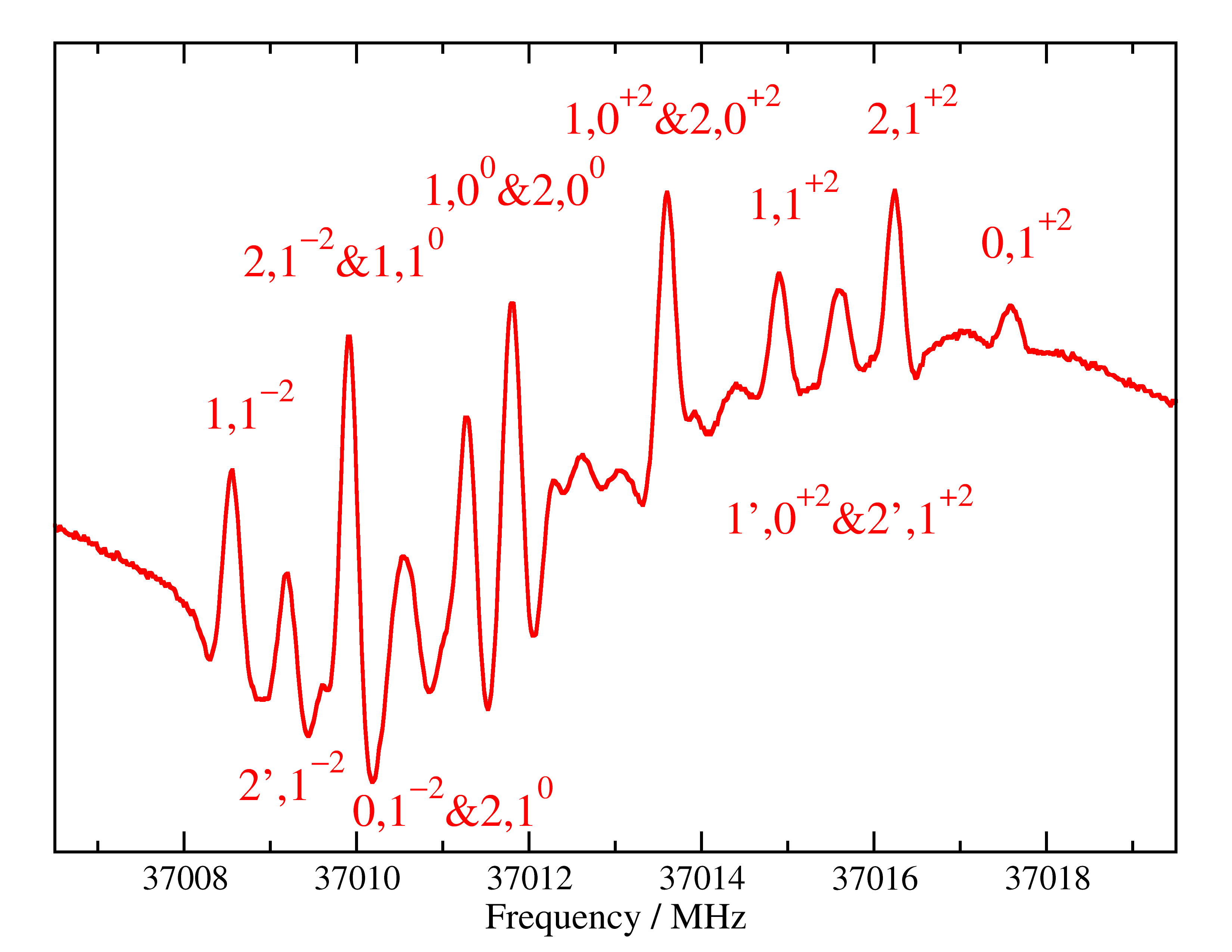}
 \end{center}
  \caption{The rotational spectrum of CH$_3$CN in the region of $\varv_8 = 2$, $J = 2 - 1$ 
           with largely resolved $^{14}$N hyperfine splitting. All but one feature (near 
           37010.5~MHz) used in the fit are indicated by assignment labels centered on 
           the respective line. The first number gives the lower state $F$ value. 
           Primed $F$ values indicate the weaker $\Delta F = 0$ transitions; unprimed 
           have $\Delta F = 1$. Numbers after the comma signal the $K$ value and, 
           as superscripts, the $l$ value. An ampersand was used to indicate overlapping 
           HFS components.}
  \label{v8_2_J1}
 \end{figure}

The rotational spectrum of methyl cyanide in its $\varv _8 = 2$ excited vibrational state 
was also studied extensively. As for $\varv _8 = 1$, the lower frequency data up to 112~GHz 
\cite{Bauer_thesis_1970,MeCN-v8=2_1969,MeCN-v8=2_1971} were in part superseded by new data 
taken in Cologne; see \textbf{Fig.~\ref{v8_2_J1}} for an overview of the $J = 2 - 1$ transition. 
The remaining data were used in the final fit with reported uncertainties. 
The $J'' = 19$ data from Ref.~\cite{MeCN-vib_le_v4_J=19_1988} had reported uncertainties 
of 50$-$150~kHz, though not specified. They fit on average to within 100~kHz and were 
retained in the final fit with assigned uncertainties of 100~kHz. The laser-sideband data 
\cite{MeCN-v8=1_2_1991} were not retained in the fit for the same reasons as in the case 
of $\varv _8 = 1$.

Interactions between $\varv _8 = 1$ and 2 were described in subsection~\ref{v8=1}.

%%%%%%%%%%%%%%%%%%%%%%%%%%%%%%%%%%%%%%%%%%%%%%%%%%%%%%%%%%%%%%%%%%%%%%%%%%%%%%%%%%%%%
%%%%%  Figure 8  %%%%%%%%%%%%%%%%%%%%%%%%%%%%%%%%%%%%%%%%%%%%%%%%%%%%%%%%%%%%%%%%%%%%
%%%%%%%%%%%%%%%%%%%%%%%%%%%%%%%%%%%%%%%%%%%%%%%%%%%%%%%%%%%%%%%%%%%%%%%%%%%%%%%%%%%%%

%%%%%%%%%%%%%%%%%%%%%%%%%%%%%%%%%%%%%%%%%%%%%%%%%%%%%%%%%%%%%%%%%%%%%%%%%%%%%%%%%%%%%
%%%%%%%%%%%%%%%%%%%%%%%%%%%%%%%%%%%%%%%%%%%%%%%%%%%%%%%%%%%%%%%%%%%%%%%%%%%%%%%%%%%%%

%%%%%%%%%%%%%%%%%%%%%%%%%%%%%%%%%%%%%%%%%%%%%%%%%%%%%%%%%%%%%%%%%%%%%%%%%%%%%%%%%%%%%
%%%%%  Figure 9  %%%%%%%%%%%%%%%%%%%%%%%%%%%%%%%%%%%%%%%%%%%%%%%%%%%%%%%%%%%%%%%%%%%%
%%%%%%%%%%%%%%%%%%%%%%%%%%%%%%%%%%%%%%%%%%%%%%%%%%%%%%%%%%%%%%%%%%%%%%%%%%%%%%%%%%%%%

 \begin{figure}
 \begin{center}
  \includegraphics[width=8.8cm]{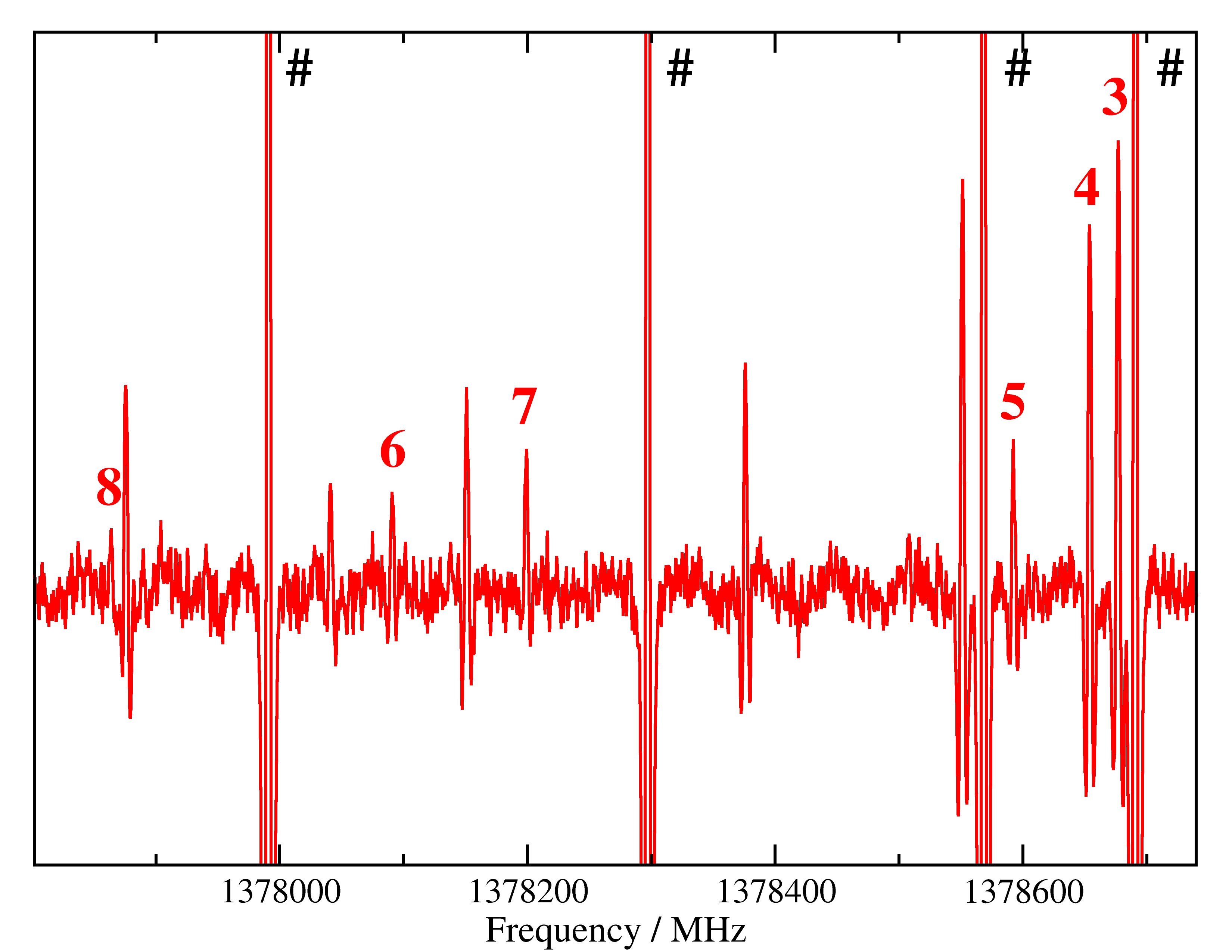}
 \end{center}
  \caption{The rotational spectrum of CH$_3$CN in the region of $J = 75 - 74$ displaying 
           the perturbation of $\varv _8 = 2^{-2}$, $K = 6$ by $\varv _4 = 1$, $K = 5$. 
           The $K$ values are given for $\varv _8 = 2^{-2}$; the perturbation shifts the 
           $K = 6$ transition lower by 350~MHz. The clipped lines marked with a black 
           pound sign are (from left to right) $K = 4$, 3, 2 and 1 transitions of 
           $\varv _8 = 1^{+1}$.}
  \label{rot_reso_v8_2_v4_1}
 \end{figure}

%%%%%%%%%%%%%%%%%%%%%%%%%%%%%%%%%%%%%%%%%%%%%%%%%%%%%%%%%%%%%%%%%%%%%%%%%%%%%%%%%%%%%
%%%%%%%%%%%%%%%%%%%%%%%%%%%%%%%%%%%%%%%%%%%%%%%%%%%%%%%%%%%%%%%%%%%%%%%%%%%%%%%%%%%%%

The energy difference between $\varv _8 = 2^0$ and $\varv _4 = 1$ is 203.54~cm$^{-1}$,  
rather large for a resonant Fermi interaction to occur. A $\Delta K = \pm1$, 
$\Delta l = \mp2$ interaction between $K = 6$ of $\varv _8 = 2^{-2}$ and $K = 5$ of 
$\varv _4 = 1$ was considered in the analysis of the $\nu _4$ band 
\cite{MeCN_nu4_nu7_3nu8_1993}. We determined a resonance to occur at $J = 73$. 
Our closest $\varv _8 = 2$ transition is $J = 75 - 74$ which is calculated 
to be shifted by $\sim$350~MHz, see \textbf{Fig.~\ref{rot_reso_v8_2_v4_1}}. 
A $\Delta K = \pm3$ ($K = 8/5$), $\Delta l = 0$ interaction, resonant at $J = 57$, 
was identified here for the first time. Our closest $\varv _8 = 2$ transitions are 
$J = 56 - 55$ and $J = 59 - 58$ with calculated shifts of about 20~MHz and 25~MHz, 
respectively.

Even though the energy difference between $\varv _8 = 2^2$ and $\varv _7 = 1$ is 
302.71~cm$^{-1}$, much larger than the difference between $\varv _8 = 2^0$ and 
$\varv _4 = 1$, the Coriolis interaction \textit{within} the $l$ components of 
$\varv _8 = 2^2$ and $\varv _7 = 1$ lead to a $\Delta K = 0$, $\Delta l = 3$ Fermi 
resonance at $K = 13$ of $\varv _8 = 2^{-2}$ and $\varv _7 = 1^{+1}$, which caused 
perturbations in the $\nu _7$ band \cite{MeCN_nu4_nu7_3nu8_1993}. We were able 
to assign transitions up to $K = 11$ in $\varv _8 = 2^{-2}$. Effects of this Fermi 
resonance were clearly present, but they were not sufficient to separate them reliably 
from the stronger Fermi resonance of $\varv _8 = 2^{-2}$ with $\varv _8 = 3^{+1}$ 
occuring at $K = 12$.

A $\Delta K = \pm2$, $\Delta l = \mp1$ resonant interaction between $K = 9$ of 
$\varv _8 = 2^{-2}$ and $K = 7$ of $\varv _7 = 1^{-1}$ near $J = 88$ shows no effects 
in the spectrum up to $J = 65$. At higher $J$, the intensities of these high energy 
lines are just too small to be identified reliably in our present spectral recordings. 
An identical type of interaction may occur at $K = 14$ and 15 of $\varv _8 = 2^0$ and 
$K = 12$ and 13 of $\varv _7 = 1^{+1}$, however, we have no assignments yet for such 
high energy transitions.

Fermi interactions between $\varv _8 = 1$ and 2 may also occur between $\varv _8 = 2$ and 3. 
There are $\Delta \varv_8 = \pm 1$, $\Delta K = 0$, $\Delta l = \pm3$ Fermi resonances 
between $\varv _8 = 2^{-2}$ and $\varv _8 = 3^{+1}$ at $K = 13$ and between $\varv _8 = 2^0$ 
and $\varv _8 = 3^{+3}$ at $K = 14/15$; a resonant interaction may occur for the latter 
at $K = 15$, currently predicted at $J = 71/72$. In the case of $\varv _8 = 2^{-2}$, 
our assignments extend to $K = 11$ and to $K = 13$ for $\varv _8 = 2^0$. These levels are 
perturbed sufficiently, up to more than 20~MHz and almost 30~MHz, respectively, 
to evaluate the corresponding Fermi parameters.

$\Delta \varv_8 = \pm 1$, $\Delta K = \mp 2$, $\Delta l = \pm 1$ interactions are 
expected to occur between $\varv _8 = 2$ and 3 at $K = 12$, $l = -2$ and $K = 10$, 
$l = -1$; at $K = 14$, $l = 0$ and $K = 12$, $l = +1$; and at $K = 16$, $l = +2$ and 
$K = 14$, $l = +3$. One tentative assignment exists for the latter interaction. 
In addition, we have 6 assignments extending to as high as $K = 19$ of 
$\varv _8 = 2^{+2}$, but several have residuals slightly larger than the estimated 
uncertainties, which may be explained by unregnonized overlap with other transitions, by 
perturbations of one or both rotational levels, or by underestimation of the uncertainties.

We should point out that there may be interactions even higher in $\Delta K$ or 
$\Delta l$, but we do not have evidence or data for these at present. 
No attempts were made to locate interations of $\varv _8 = 2$ with vibrational states 
higher than $\varv _8 = 3$. Such interactions may be non-negligible, but most likely 
at $K$ levels with few or no assignments in our present data set.

%%%%%%%%%%%%%%%%%%%%%%%%%%%%%%%%%%%%%%%%%%%%%%%%%%%%%%%%%%%%%%%%%%%%%%%%%%%%%%%%%%%%%
\subsubsection{The 2$\nu_8$ band}
%%%%%%%%%%%%%%%%%%%%%%%%%%%%%%%%%%%%%%%%%%%%%%%%%%%%%%%%%%%%%%%%%%%%%%%%%%%%%%%%%%%%%

Transition frequencies were determined entirely from our new IR spectrum since 
Ref.~\cite{MeCN_2nu8_1993} did not report their experimental values.

The strongly allowed transitions in the $2\nu_8$ IR band are those with 
$\Delta K = \Delta l = 0$. The $K$ quantum numbers reach 12, higher by 2 than 
in the previous study \cite{MeCN_2nu8_1993}, while the maximum value of $J$ is 
essentially identical, 66 versus 65. As in the previous study, we were unable to make 
any assignments in the perpendicular $\Delta l = \pm2$, $\Delta K = \mp1$ component, 
which is only weakly allowed, but we were able to make numerous assignments to levels 
in the $l = \pm2$ component which gain intensity through $\Delta K = \Delta l = 2$ 
mixing. This mixing is particularly strong for $K = 4$ of $\varv _8 = 2^0$ and 
$K = 2$ of $\varv _8 = 2^{-2}$ with the former being lower in energy at low-$J$ 
until the energy ordering changes at $J = 40/41$. 
As a consequence, many $^oR_4(J)$ and $^oP_4(J)$ IR transitions were asssigned 
covering most of the $J$ from 8 to 62. It is interesting to note that the mixing 
of these $K$ levels changes only slightly between adjacent $J$ levels such that 
rotational transitions between these mixed $K$ levels are too weak to be observed 
with the exception of those at the change of the energy ordering. 
As also shown in \textbf{Table~\ref{statistics}}, the IR assignments to 
$\varv _8 = 2^{-2}$ cover mostly $K = 1$ to 3 with several $K = 4$ assignments and 
few higher ones, those to $\varv _8 = 2^{+2}$ cover mostly $K = 0$ to 2; a few 
sparse higher-$K$ assignments are viewed as tentative and were weighted out 
in the final fit.

Isolated lines with sufficiently good signal-to-noise ratio were given uncertainties 
of 0.0002~cm$^{-1}$, as in the case of $\nu _8$. Noisier lines or lines  in more 
crowded regions, such as in the $Q$-branch, were given larger uncertainties of 
0.0003~cm$^{-1}$ to 0.0010~cm$^{-1}$.

We were also able to make assignments in the $3\nu_8 - \nu_8$ hot band and in the 
$\nu_7 - \nu_8$ difference band, as was done earlier \cite{MeCN_2nu8_1993}. The latter 
band gains intensity because of an anharmonic resonance between $\varv _7 = 1$ and 
$\varv _8 = 3$ \cite{MeCN_2nu8_1993}. We will publish our detailed account on these 
states separately. Judging from the observation of transitions in $2\nu_8^0$ with 
$K = 12$, we may be even able to make assignments in $4\nu_8 - 2\nu_8$.

%%%%%%%%%%%%%%%%%%%%%%%%%%%%%%%%%%%%%%%%%%%%%%%%%%%%%%%%%%%%%%%%%%%%%%%%%%%%%%%%%%%%%
\subsubsection{The $2\nu_8 - \nu_8$ hot band}
%%%%%%%%%%%%%%%%%%%%%%%%%%%%%%%%%%%%%%%%%%%%%%%%%%%%%%%%%%%%%%%%%%%%%%%%%%%%%%%%%%%%%
\label{hot-band-results}

Because of the strength of the $2\nu _8^0$ band, little can be gained from the analysis 
of the $2\nu _8^0 - \nu _8$ hot band. On the other hand, IR transitions to $2\nu _8^2$ 
are limited to those which gain intensity through $q_{22}$ interaction, see previous 
subsubsection. These transitions belong to the $l = -2$ component for the most part. 
According to Koivusaari et al. \cite{MeCN_2nu8_1993}, transitions of $2\nu _8^2 - \nu _8$ 
are frequently overlapping each other or are blended with cold band transitions. 
Nevertheless, they made 52 $^rR$-branch assignments with $3 \le K'' \le 7$. 50 of these 
assignments were made in the $l = +2$ component of $2\nu _8^2 - \nu _8$ of which 6 had 
zero weight. Since no IR transition frequencies were published in Ref.~\cite{MeCN_2nu8_1993}, 
we searched our $\nu _8$ spectra for suitable hot band transitions. Unfortunately, the 
congestion of the spectral lines and the quality of the spectra were such that only 
few lines could be assigned with certainty.

%%%%%%%%%%%%%%%%%%%%%%%%%%%%%%%%%%%%%%%%%%%%%%%%%%%%%%%%%%%%%%%%%%%%%%%%%%%%%%%%%%%%%
%%%%%%%%%%%%%%%%%%%%%%%%%%%%%%%%%%%%%%%%%%%%%%%%%%%%%%%%%%%%%%%%%%%%%%%%%%%%%%%%%%%%%
%%%%%   Table 3  %%%%%%%%%%%%%%%%%%%%%%%%%%%%%%%%%%%%%%%%%%%%%%%%%%%%%%%%%%%%%%%%%%%%
%%%%%%%%%%%%%%%%%%%%%%%%%%%%%%%%%%%%%%%%%%%%%%%%%%%%%%%%%%%%%%%%%%%%%%%%%%%%%%%%%%%%%
%%%%%%%%%%%%%%%%%%%%%%%%%%%%%%%%%%%%%%%%%%%%%%%%%%%%%%%%%%%%%%%%%%%%%%%%%%%%%%%%%%%%%

\begin{table*}
\begin{center}
\caption{Spectroscopic parameters or differences $\Delta$ thereof $^{a,b}$ (cm$^{-1}$, MHz)$^c$ of 
         methyl cyanide within vibrational states $\varv_8 \le 2$.}
\label{parameter_v8_le_2}
% \smallskip
{\footnotesize
\begin{tabular}[t]{lr@{}lr@{}lr@{}lr@{}l}
\hline 
Parameter $X$ & \multicolumn{2}{c}{$\varv = 0$} & \multicolumn{2}{c}{$\varv_8 = 1$} &
 \multicolumn{2}{c}{$\varv_8 = 2^0$}  & \multicolumn{2}{c}{$\varv_8 = 2^2$}  \\
\hline
$E_{\rm{vib}}$$^c$                          &        0&.0              &      365&.024\,365\,(9)     &    716&.750\,42\,(13)     &      739&.148\,225\,(56)      \\
$(\Delta)^b\ (A - B)$                       & 148\,900&.103\,(66)      &   $-$115&.930\,(26)         & $-$187&.404\,(18)         &   $-$259&.956\,(122)          \\
$(\Delta)^b\ B$                             &   9\,198&.899\,167\,(11) &       27&.530\,277\,(49)    &     54&.057\,316\,(111)   &       54&.502\,729\,(70)      \\
$(\Delta)^b\ D_K \times 10^3$               &   2\,830&.6\,(18)        &    $-$11&.46\,(48)          &  $-$20&.19\,(34)          &     $-$7&.45\,(164)           \\
$(\Delta)^b\ D_{JK} \times 10^3$            &      177&.407\,87\,(25)  &        0&.987\,48\,(60)     &      1&.675\,5\,(25)      &        1&.808\,8\,(15)        \\
$(\Delta)^b\ D_J \times 10^6$               &   3\,807&.576\,(8)       &       95&.599\,(17)         &    216&.319\,(37)         &      189&.162\,(31)           \\
$(\Delta)^b\ H_K \times 10^6$               &      164&.6\,(66)        &       14&.9\,(22)           &       &                   &         &                     \\
$(\Delta)^b\ H_{KJ} \times 10^6$            &        6&.062\,0\,(14)   &        0&.034\,1\,(22)      &      0&.150\,3\,(239)     &        0&.024\,9\,(55)        \\
$(\Delta)^b\ H_{JK} \times 10^9$            &   1\,025&.69\,(15)       &        2&.59\,(6)           &     17&.71\,(34)          &     $-$0&.37\,(21)            \\
$(\Delta)^b\ H_J \times 10^{12}$            &   $-$237&.4\,(21)        &      315&.3\,(30)           &    200&.7\,(63)           &      627&.8\,(59)             \\
$(\Delta)^b\ L_{KKJ} \times 10^{12}$        &   $-$444&.3\,(25)        &         &                   &       &                   &         &                     \\
$(\Delta)^b\ L_{JK} \times 10^{12}$         &    $-$52&.75\,(51)       &         &                   &       &                   &         &                     \\
$(\Delta)^b\ L_{JJK} \times 10^{12}$        &     $-$7&.901\,(32)      &         &                   &       &                   &         &                     \\
$(\Delta)^b\ L_J \times 10^{15}$            &     $-$3&.10\,(17)       &     $-$2&.64\,(20)$^d$      &   $-$5&.28\,(40)$^d$      &     $-$5&.28\,(40)$^d$        \\
$(\Delta)^b\ P_{JK} \times 10^{18}$         &      552&.\,(68)         &         &                   &       &                   &         &                     \\
$(\Delta)^b\ P_{JJK} \times 10^{18}$        &       55&.3\,(22)        &         &                   &       &                   &         &                     \\
$(\Delta)^b\ eQq$                           &     $-$4&.223\,08\,(107) &     $-$0&.038\,7\,(19)$^d$  &   $-$0&.077\,4\,(38)$^d$  &     $-$0&.077\,4\,(38)$^d$    \\
$eQq\eta$                                   &         &                &        0&.1519\,(113)       &       &                   &         &                     \\
$(\Delta)^b\ C_{bb} \times 10^3$            &        1&.845\,(90)      &         &                   &       &                   &         &                     \\
$(\Delta)^b\ (C_{aa} - C_{bb}) \times 10^3$ &     $-$1&.15\,(31)       &         &                   &       &                   &         &                     \\
$A\zeta$                                    &         &                & 138\,656&.195\,(73)         &       &                   & 138\,656&.042\,(102)          \\
$\eta_K$                                    &         &                &       10&.332\,9\,(72)      &       &                   &       10&.405\,1\,(98)        \\
$\eta_J$                                    &         &                &        0&.390\,469\,(7)     &       &                   &        0&.394\,512\,(6)       \\
$\eta_{KK} \times 10^6$                     &         &                &   $-$834&.\,(41)$^d$        &       &                   &   $-$834&.\,(41)$^d$          \\
$\eta_{JK} \times 10^6$                     &         &                &    $-$34&.057\,(64)$^d$     &       &                   &    $-$34&.057\,(64)$^d$       \\
$\eta_{JJ} \times 10^6$                     &         &                &     $-$2&.359\,5\,(24)$^d$  &       &                   &     $-$2&.359\,5\,(24)$^d$    \\
$\eta_{JKK} \times 10^9$                    &         &                &        2&.59\,(17)$^d$      &       &                   &        2&.59\,(17)$^d$        \\
$\eta_{JJK} \times 10^9$                    &         &                &        0&.508\,5\,(63)$^d$  &       &                   &        0&.508\,5\,(63)$^d$    \\
$q$                                         &         &                &       17&.798\,438\,(23)    &       &$^e$               &       17&.729\,857\,(138)$^e$ \\
$q_{K} \times 10^3$                         &         &                &     $-$2&.664\,5\,(111)     &       &$^e$               &     $-$2&.615\,3\,(89)$^{e}$  \\
$q_{J} \times 10^6$                         &         &                &    $-$63&.842\,(14)         &       &$^e$               &    $-$68&.668\,(31)$^e$       \\
$q_{JK} \times 10^9$                        &         &                &       93&.19\,(53)$^d$      &       &$^e$               &       93&.19\,(53)$^{d,e}$    \\
$q_{JJ} \times 10^{12}$                     &         &                &      311&.5\,(15)           &       &$^e$               &      191&.9\,(26)$^e$         \\
\hline \hline
\end{tabular}\\[2pt]
}
\end{center}
{\footnotesize
$^a$ Numbers in parentheses are one standard deviation in units of the least significant figures. 
     Empty entries indicate parameter not applicable or not used in the fit. See section~\ref{det_spec_parameters} for sign and value considerations.\\
$^b$ Parameter $X$ given for $\varv = 0$; $\Delta X = X_i - X_0$, with $i$ representing an excited 
     vibrational state.\\
$^c$ All parameters given in units of megahertz, except for $E_{\rm{vib}}$, which is given in 
     units of inverse centimeters.\\
$^d$ Ratio kept fixed in the fit for respective $X$ or $\Delta X$.\\
$^e$ The parameter $q$ and its distortion corrections connect levels with $\Delta K = \Delta l = 2$, 
     see subsection~\ref{intro-spec} and section~\ref{det_spec_parameters}.\\
}
\end{table*}

%%%%%%%%%%%%%%%%%%%%%%%%%%%%%%%%%%%%%%%%%%%%%%%%%%%%%%%%%%%%%%%%%%%%%%%%%%%%%%%%%%%%%
%%%%%%%%%%%%%%%%%%%%%%%%%%%%%%%%%%%%%%%%%%%%%%%%%%%%%%%%%%%%%%%%%%%%%%%%%%%%%%%%%%%%%

%%%%%%%%%%%%%%%%%%%%%%%%%%%%%%%%%%%%%%%%%%%%%%%%%%%%%%%%%%%%%%%%%%%%%%%%%%%%%%%%%%%%%
\subsection{Remarks on higher vibrational states}
%%%%%%%%%%%%%%%%%%%%%%%%%%%%%%%%%%%%%%%%%%%%%%%%%%%%%%%%%%%%%%%%%%%%%%%%%%%%%%%%%%%%%

The vibrational state $\varv _8 = 2$ interacts with $\varv _4 = 1$, with $\varv _7 = 1$, 
and with $\varv _8 = 3$. Therefore, we need to take into account previously determined 
parameters for these states \cite{MeCN_nu4_nu7_3nu8_1993}, which involve interactions 
among these three vibrational states.

There is a $\Delta K = \Delta l = \pm1$ Coriolis interaction between $\varv _4 = 1$ and 
$\varv _7 = 1$ which will be resonant at $K = 25$ or slightly higher for $v _7 = 1^{-1}$, 
too high to be observed easily. Another interaction occurs between $K = 6/7$ of 
$\varv _4 = 1$ and $K = 4/5$ of $\varv _7 = 1^{+1}$.

An anharmonic resonance occurs at $K = 7/8$ of $\varv _4 = 1$ and $\varv _8 = 3^{+3}$.
Another, stronger anharmonic resonance occurs at $K = 7/8$ of $\varv _7 = 1^{+1}$ and 
$\varv _8 = 3^{+1}$. Inspection of preliminary energy levels involving these states 
suggest a plethora of further interactions among these states. Such interactions 
are also possible with states of the next polyad consisting of $\varv _4 = \varv _8 = 1$, 
$\varv _3 = 1$, $\varv _7 = \varv _8 = 1$, $\varv _6 = 1$, and $\varv _8 = 4$ 
\cite{pentade_1994}.

%%%%%%%%%%%%%%%%%%%%%%%%%%%%%%%%%%%%%%%%%%%%%%%%%%%%%%%%%%%%%%%%%%%%%%%%%%%%%%%%%%%%%
%%%%%%%%%%%%%%%%%%%%%%%%%%%%%%%%%%%%%%%%%%%%%%%%%%%%%%%%%%%%%%%%%%%%%%%%%%%%%%%%%%%%%
%%%%%  Determination of spectroscopic parameters  %%%%%%%%%%%%%%%%%%%%%%%%%%%%%%%%%%%
%%%%%%%%%%%%%%%%%%%%%%%%%%%%%%%%%%%%%%%%%%%%%%%%%%%%%%%%%%%%%%%%%%%%%%%%%%%%%%%%%%%%%
%%%%%%%%%%%%%%%%%%%%%%%%%%%%%%%%%%%%%%%%%%%%%%%%%%%%%%%%%%%%%%%%%%%%%%%%%%%%%%%%%%%%%
\section{Determination of spectroscopic parameters}
\label{det_spec_parameters}

Pickett's {\scriptsize SPCAT} and {\scriptsize SPFIT} programs \cite{spfit_1991} were used for 
prediction of the CH$_3$CN spectra and for fitting of the measured data. The programs were 
intended to be rather general, thus being able to fit asymmetric top rotors with spin and 
vibration-rotation interaction. They have evolved considerably with time because many aspects 
were not available initially \cite{editorial_Herb-Ed,intro_JPL-catalog}, 
in particular special considerations for symmetric or linear molecules or for higher 
symmetry cases. Spectra of symmetric top rotors with excitation of one 
\cite{MeCCH_nu10+Dyade_2004,H3O+_rot_fitting_2009} or two quanta of a degenerate 
vibrational mode \cite{MeCCH_nu10+Dyade_2004} were fit successfully with 
{\scriptsize SPFIT}. The $l$-doubling is implemented as $l \equiv 0, \pm1$~mod~3, thus 
$l = \pm2$ are identified with $l = \mp1$. Hence, $G_a$, the $\Delta K = 0$ Coriolis 
interaction parameter between the $l$-components of $\varv_8 = 1^1$ and $\varv_8 = 2^2$ 
equals $-2A\zeta$ and $+4A\zeta$, respectively. In fact, the Coriolis term enters the 
Hamiltonian as $-2(A\zeta)_{\varv}kl$, see, e.g., Eq.~(1) in Refs.~\cite{MeCN_nu8_1992} 
and \cite{MeCN_2nu8_1993}. The $J$- and $K$-dependent distortion parameters to $G_a$ 
are usually defined without the factor of $-2$, hence $G_a^J$ equals $\eta_J$ and 
$-2\eta_J$ for $\varv_8 = 1$ and $\varv_8 = 2$, respectively, and so on. 
Similarly, e.g., $G_a$ and $F_{ab}$, are the coefficients of $iJ_a$ and 
$(J_aJ_b + J_bJ_a)/2$, respectively. When they act as Coriolis parameters 
between two interacting states, the values from {\scriptsize SPFIT} often equal 
twice the values from other fitting programs. 
This is likely a consequence of the definition of $G_i$ which is approximately 
equal to $2B_i\zeta _i$. $F$ designates a Fermi (more generally anharmonic) 
interaction parameter. $J$ or $K$ as sub- or superscripts represent, as usual, 
corresponding distortion corrections. $F_2$, $G_{2i}$, and $F_{2ij}$ are 
the coefficients of operators which connect levels with $\Delta K$ higher by 2 than 
$F$, $G_{i}$, and $F_{ij}$, respectively, and with $\Delta l$ modified accordingly 
to meet the symmetry condition $\Delta K - \Delta l \equiv 0$~mod~3. I.e. $X_2$ is 
defined for a parameter $X$ as coefficient of $\{{\bf X},{\bf J_+}^2 {\bf L_+}^2 + 
{\bf J_-}^2 {\bf L_-}^2\}$, with $\{a,b\} = (ab +ba)/2$ being the anticommutator. 
Interaction parameters between different vibrations are indicated by numbers in 
parentheses separated by a comma; the degree of excitation and the $l$ quantum numbers 
are also given if necessary. The most important interaction parameter with $\Delta K > 0$ 
within a degenerate vibrational state of a strongly prolate symmetric rotor is $q$, also 
known as $q_{22}$ because it connects levels differing in $\Delta K = \Delta l = 2$. 
It can have large effects in particular in low-lying bending modes, in which the 
magnitude of $\zeta$ often is close to the limiting case of 1.0, which will cause 
ample near-degeneracies. Scaling factors for $q$ depending on the $l$-components and 
the vibrational excitation are not implemented in {\scriptsize SPCAT} and 
{\scriptsize SPFIT}. The values are $q/2$ and $q/\sqrt2$ for $\varv _8 = 1$ and 2, 
respectively. We note that $q$ and its distortion corrections are sometimes defined 
negatively. This applies to the studies on methyl cyanide  
\cite{MeCN_nu8_1992,MeCN_2nu8_1993,MeCN_nu4_nu7_3nu8_1993,pentade_1994} and propyne 
\cite{MeCCH_nu10+Dyade_2002,MeCCH_nu10+Dyade_2004,MeCCH_Dyade+Triade_2009}. Therefore, 
we inverted the signs and added footnotes to the appropriate tables. We also note an 
additional factor of 2 in the definitions of $q_K$, $q_{JK}$, etc. in the articles on 
propyne, causing the determined numerical values to be smaller than in {\scriptsize SPFIT}. 
However, none of these parameters were used in the IR studies of methyl cyanide relevant 
to our investigation \cite{MeCN_nu8_1992,MeCN_2nu8_1993,MeCN_nu4_nu7_3nu8_1993,pentade_1994}.

%%%%%%%%%%%%%%%%%%%%%%%%%%%%%%%%%%%%%%%%%%%%%%%%%%%%%%%%%%%%%%%%%%%%%%%%%%%%%%%%%%%%%
%%%%%   Table 4  %%%%%%%%%%%%%%%%%%%%%%%%%%%%%%%%%%%%%%%%%%%%%%%%%%%%%%%%%%%%%%%%%%%%
%%%%%%%%%%%%%%%%%%%%%%%%%%%%%%%%%%%%%%%%%%%%%%%%%%%%%%%%%%%%%%%%%%%%%%%%%%%%%%%%%%%%%

\begin{table*}
\begin{center}
\caption{Spectroscopic parameters or differences $\Delta$ thereof $^{a,b}$ (cm$^{-1}$, MHz)$^c$ of 
         methyl cyanide within vibrational states $\varv_4 = 1$, $\varv_7 = 1$, and $\varv_8 = 3$ 
         taken from Ref.~\cite{MeCN_nu4_nu7_3nu8_1993}.}
\label{parameter_triade}
% \smallskip
{\footnotesize
\begin{tabular}[t]{lr@{}lr@{}lr@{}lr@{}l}
\hline 
Parameter $X$ & \multicolumn{2}{c}{$\varv_4 = 1$} & \multicolumn{2}{c}{$\varv_7 = 1$} &
 \multicolumn{2}{c}{$\varv_8 = 3^1$}  & \multicolumn{2}{c}{$\varv_8 = 3^3$}  \\
\hline
$E_{\rm{vib}}$$^c$           &    920&.290\,03  &   1\,041&.854\,71     &   1\,077&.786\,3      &   1\,122&.15          \\
$\Delta (A - B)$             & $-$166&.205      &      889&.440         &   $-$302&.030         &   $-$347&.76          \\
$\Delta B$                   &  $-$46&.148\,22  &     $-$5&.734\,13     &       80&.294\,85     &       81&.289         \\
$\Delta D_K \times 10^3$     &  $-$31&.63       &      149&.333         &    $-$22&.8$^d$       &    $-$22&.8$^d$       \\
$\Delta D_{JK} \times 10^3$  &      7&.165      &        0&.973\,1      &        3&.504\,6$^d$  &        3&.504\,6$^d$  \\
$\Delta D_J \times 10^6$     &   $-$5&.702\,05  &       14&.001         &      462&.$^d$        &      462&.$^d$        \\
$\Delta H_K \times 10^6$     &  $-$22&.24       &         &             &         &             &         &             \\
$\Delta H_{KJ} \times 10^6$  &  $-$13&.895      &         &             &         &             &         &             \\
$\Delta H_{JK} \times 10^9$  &    315&.68       &   $-$148&.1           &         &             &         &             \\
$A\zeta$                     &       &          &  66\,663&.668         & 138\,665&.87          & 138\,527&.8           \\
$\eta_K$                     &       &          &        7&.224\,8      &       11&.013$^d$     &       11&.013$^d$     \\
$\eta_J$                     &       &          &        0&.078\,153    &        0&.401\,54$^d$ &        0&.401\,54$^d$ \\
$\eta_{KK} \times 10^6$      &       &          &       64&.8           &         &             &         &             \\
$\eta_{JK} \times 10^6$      &       &          &    $-$50&.34          &         &             &         &             \\
$\eta_{JJ} \times 10^6$      &       &          &        2&.385         &        7&.285$^d$     &        7&.285$^d$     \\
$q$                          &       &          &        4&.763\,4      &       17&.683$^{d,e}$ &       17&.683$^{d,e}$ \\
$q_{J} \times 10^6$          &       &          &    $-$10&.85          &    $-$75&.79$^{d,e}$  &    $-$75&.79$^{d,e}$  \\
\hline \hline
\end{tabular}\\[2pt]
}
\end{center}
{\footnotesize
$^a$ Empty entries indicate parameter not applicable or not used in the fit. 
     Signs of $q$ and $q_J$ altered, see section~\ref{det_spec_parameters}.\\
$^b$ Parameter difference $\Delta (X) = X_i - X_0$, given for rotational and distortion parameters  
     ($i$ represents an excited vibrational state).\\
$^c$ All parameters given in units of megahertz, except for $E_{\rm{vib}}$, which is given in 
     units of inverse centimeters.\\
$^d$ Ratio kept fixed in the fit for parameter $X$ or $\Delta X$.\\
$^e$ The parameter $q$ and its distortion corrections connect levels with $\Delta K = \Delta l = 2$, 
     see subsection~\ref{intro-spec} and section~\ref{det_spec_parameters}.\\
}
\end{table*}

%%%%%%%%%%%%%%%%%%%%%%%%%%%%%%%%%%%%%%%%%%%%%%%%%%%%%%%%%%%%%%%%%%%%%%%%%%%%%%%%%%%%%
%%%%%%%%%%%%%%%%%%%%%%%%%%%%%%%%%%%%%%%%%%%%%%%%%%%%%%%%%%%%%%%%%%%%%%%%%%%%%%%%%%%%%

We determined rotational, centrifugal, and hyperfine structure (HFS) parameters of the ground 
state as common for all vibrational states. As most of the data, in particular the IR data, 
were not affected by HFS, all states were defined twice, with and without HFS. Vibrational 
changes $\Delta X = X_{\rm i} - X_0$, where $X$ represents a parameter and $X_{\rm i}$ the 
parameter in an excited vibrational state. This is very similar to several previous studies 
on CH$_3$CN \cite{MeCN_nu8_1992,MeCN_2nu8_1993,MeCN_nu4_nu7_3nu8_1993,pentade_1994} 
and rather convenient because vibrational corrections $\Delta X$ are usually small 
with respect to $X$ especially for lower order parameters $X$. In addition, this offers 
the opportunity to constrain vibrational corrections to $\varv_8 = 2$ to twice those of 
$\varv_8 = 1$, thus reducing the amount of independent spectroscopic parameters further.

The degenerate vibrational states required $A\zeta$ and $q$ to be used in the fit with 
centrifugal distortion corrections as far as needed. Several of the distortion corrections 
between $\varv_8 = 1$ and 2 were constrained to be identical. Fitting of our new low 
frequency data of $\varv_8 = 1$ required $eQq\eta$ to be used in the fit, the quadrupole 
parameter equivalent to $q$. We assumed that the value determined by {\scriptsize SPFIT} 
is $+eQq\eta/4$

Starting values for the parameters of $\varv = 0$, $\varv_8 = 1$, and $\varv_8 = 2$, were 
taken from Refs.~\cite{MeCN_rot_2009,MeCN_nu8_1992,MeCN_2nu8_1993}. Parameters describing 
the triad around 1000~cm$^{-1}$, $\varv_4 = 1$, $\varv_7 = 1$, and $\varv_8 = 3$, were 
taken from Ref.~\cite{MeCN_nu4_nu7_3nu8_1993} and were kept fixed with the exception of 
one rotational interaction parameter between $\varv_8 = 2$ and $\varv_4 = 1$.

Parameters were introduced newly to the fit, or a constraint was released, if the rms error 
(as the measure of the quality of the fit) was reduced by an amount deemed significant 
and if the new parameter was useful in the Hamiltonian and if the resulting value(s) 
appeared to be reasonable. The latter aspect was a difficult one because vibrational 
corrections to some higher order parameters were of similar magnitudes as the parameters 
themselves and because some vibrational changes to $\varv_8 = 2^0$ or $\varv_8 = 2^2$ were 
rather different from twice the corresponding vibrational change to $\varv_8 = 1$. 
In most instances, care was taken to search for the largest reduction of the rms error. 
In almost all instances, this led to parameters determined with high significance, 
i.e. their values were more than five times larger in magnitude than their uncertainties. 
In few cases, parameters not determined with high significance were kept in the fit 
because the uncertainties were of similar magnitudes as those for other vibrational states.

Concerning the interactions between $\varv _8 = 1$ and 2, we determined $J$ distortion 
corrections $F_J$ and $F_{JJ}$ to the main Fermi term $F$, whereas $F_K$ was not determined 
with significance. We estimated its value by assuming $F_K/F_J \approx A/B$. 
Once both $F_2$ parameters were well-determined, their ratio was essentially 2 within the 
uncertainties. Therefore, the ratio was constrained to reduce the number of independent 
parameters. No distortion corrections could be determined for the Fermi terms associated 
with the interactions between $\varv _8 = 2$ and 3, so the $J$ and $K$ distortion 
corrections were fixed to appropriate multiples of the values determined for the 
$\varv _8 = 1$/2 resonance. Trial fits with $F_2$ yielded values or constraints which 
were in some cases much smaller than those in the $\varv _8 = 1$/2 resonances with 
essentially no effect on the quality on the fit. These parameters were omitted from 
the final fits.

The parameter $f_{44}$ (for $\varv _8 = 2$) had little effects on the rms error in 
most trial fits and was not determined with significance. It seemed to be determined 
in some trial fits, but its magnitude appeared to be too large with respect to that 
of $\varv _{10} = 2$ of propyne \cite{MeCCH_nu10+Dyade_2004}. Moreover, changes in 
some of the remaining spectroscopic parameters were deemed too large in these cases. 
Therefore, $f_{44}$ was omitted from the final fit.

%%%%%%%%%%%%%%%%%%%%%%%%%%%%%%%%%%%%%%%%%%%%%%%%%%%%%%%%%%%%%%%%%%%%%%%%%%%%%%%%%%%%%
%%%%%   Table 5  %%%%%%%%%%%%%%%%%%%%%%%%%%%%%%%%%%%%%%%%%%%%%%%%%%%%%%%%%%%%%%%%%%%%
%%%%%%%%%%%%%%%%%%%%%%%%%%%%%%%%%%%%%%%%%%%%%%%%%%%%%%%%%%%%%%%%%%%%%%%%%%%%%%%%%%%%%

\begin{table}
\begin{center}
\caption{Interaction parameters$^{a}$ (MHz) between vibrational states of methyl cyanide.}
\label{interation-parameter}
% \smallskip
{\footnotesize
 \renewcommand{\arraystretch}{1.10}
\begin{tabular}[t]{lr@{}l}
\hline 
Parameter                                    & \multicolumn{2}{c}{Value}    \\
\hline
$F_2(0,8^1) \times 10^3$                     &     $-$70&.897\,(27)         \\
$F(8^{\pm1},8^{2,\mp2})$                     &   53\,157&.7\,(33)           \\
$F_K(8^{\pm1},8^{2,\mp2})$                   &      $-$6&.$^{b,c}$          \\
$F_J(8^{\pm1},8^{2,\mp2}) \times 10^3$       &    $-$369&.89\,(44)$^b$      \\
$F_{JJ}(8^{\pm1},8^{2,\mp2}) \times 10^6$    &         1&.681\,(87)$^b$     \\
$F_2(8^{\pm1},8^{2,0}) \times 10^3$          &     $-$65&.491\,(24)$^d$     \\
$F_2(8^{\pm1},8^{2,\pm2}) \times 10^3$       &    $-$130&.982\,(48)$^d$     \\
$F_{ac}(8^{2,\pm2},4^1)$ $[2w_{488}]$        &         8&.736\,2\,(21)      \\
$F_{2ac}(8^{2,0},4^1) \times 10^6$           &      $-$7&.98\,(16)          \\
$F(8^{\pm2},7^{\mp1})$ $[W_{788}]$           &   45\,170&.8$^e$             \\
$F(8^{2,\pm2},8^{3,\mp1})$                   &   77\,208&.\,(93)            \\
$F(8^{2,0},8^{3,3})$                         &   91\,509&.\,(131)           \\
$G_b(4,7)$ $[2W_{47}]$                       &       909&.$^e$              \\
$F_{bc}(4,7)$ $[2w_{47}]$                    &      $-$1&.84$^e$            \\
$F(4,8^{3,\pm3}))$ $[W_{4888}]$              &   11\,430&.$^e$              \\
$F(7^{\pm1},8^{3,\pm1})$ $[W_{7888}]$        &   50\,129&.2$^e$             \\
$G_a(7^{\pm1},8^{3,\pm1})$ $[2W^k_{7888}]$   & $-$2\,239&.1$^e$             \\
\hline \hline 
\end{tabular}\\[2pt]
}
\end{center}
{\footnotesize
$^a$ Alternative designations from Ref.~\cite{MeCN_nu4_nu7_3nu8_1993} given in brackets. 
     Numbers in parentheses after the interaction parameter designate the vibrational 
     states separated by a comma, see also section~\ref{det_spec_parameters}. Numbers in 
     parentheses after the values are one standard deviation in units of the least 
     significant figures.\\
$^b$ $J$ and $K$ distortion corrections to $F(8^{2,\pm2},8^{3,\mp1})$ and $F(8^{2,0},8^{3,3})$ 
     kept fixed to $\sqrt2$ and $\sqrt3$, respectively, times the corresponding 
     $F(8^{\pm1},8^{2,\mp2})$ value; see, e.g., Ref.~\cite{MeCCH_nu10+Dyade_2004}.\\
$^c$ Estimated assuming $F_K/F_J \approx A/B$, see section~\ref{det_spec_parameters}.\\
$^d$ Ratio constrained, see section~\ref{det_spec_parameters}.\\
$^e$ Kept fixed to values from Ref.~\cite{MeCN_nu4_nu7_3nu8_1993}.\\
}
\end{table}

%%%%%%%%%%%%%%%%%%%%%%%%%%%%%%%%%%%%%%%%%%%%%%%%%%%%%%%%%%%%%%%%%%%%%%%%%%%%%%%%%%%%%
%%%%%%%%%%%%%%%%%%%%%%%%%%%%%%%%%%%%%%%%%%%%%%%%%%%%%%%%%%%%%%%%%%%%%%%%%%%%%%%%%%%%%
%%%%%%%%%%%%%%%%%%%%%%%%%%%%%%%%%%%%%%%%%%%%%%%%%%%%%%%%%%%%%%%%%%%%%%%%%%%%%%%%%%%%%

%%%%%%%%%%%%%%%%%%%%%%%%%%%%%%%%%%%%%%%%%%%%%%%%%%%%%%%%%%%%%%%%%%%%%%%%%%%%%%%%%%%%%
%%%%%%%%%%%%%%%%%%%%%%%%%%%%%%%%%%%%%%%%%%%%%%%%%%%%%%%%%%%%%%%%%%%%%%%%%%%%%%%%%%%%%
%%%%%   Table 6  %%%%%%%%%%%%%%%%%%%%%%%%%%%%%%%%%%%%%%%%%%%%%%%%%%%%%%%%%%%%%%%%%%%%
%%%%%%%%%%%%%%%%%%%%%%%%%%%%%%%%%%%%%%%%%%%%%%%%%%%%%%%%%%%%%%%%%%%%%%%%%%%%%%%%%%%%%
%%%%%%%%%%%%%%%%%%%%%%%%%%%%%%%%%%%%%%%%%%%%%%%%%%%%%%%%%%%%%%%%%%%%%%%%%%%%%%%%%%%%%

\begin{table*}
\begin{center}
\caption{Spectroscopic parameters or differences $\Delta$ thereof $^{a,b}$ (cm$^{-1}$, MHz)$^c$ of 
         methyl cyanide involving $\varv_8 = 1$ and 2 in comparison to previous values.}
\label{comparison_v8_le_2_new-old}
% \smallskip
{\footnotesize
\begin{tabular}[t]{lr@{}lr@{}llr@{}lr@{}llr@{}lr@{}l}
\hline 
              &  \multicolumn{4}{c}{$\varv_8 = 1$} & & \multicolumn{4}{c}{$\varv_8 = 2^0$} & & \multicolumn{4}{c}{$\varv_8 = 2^2$}  \\
\cline{2-5} \cline{7-10} \cline{12-15} 
Parameter $X$ & \multicolumn{2}{c}{Present} & \multicolumn{2}{c}{Ref.~\cite{MeCN_nu8_1992}} & & \multicolumn{2}{c}{Present} & \multicolumn{2}{c}{Ref.~\cite{MeCN_2nu8_1993}} & & 
\multicolumn{2}{c}{Present} & \multicolumn{2}{c}{Ref.~\cite{MeCN_2nu8_1993}}  \\
\hline
$E_{\rm{vib}}$$^c$          &    365&.024365~(9)    &    365&.015965~(12) & &    716&.75042~(13)  &    716&.70441~(1)  & &    739&.14823~(6)       &    739&.12907~(3)      \\
$\Delta A$                  &  $-$88&.400~(26)      &  $-$90&.079~(36)    & & $-$133&.347~(18)    & $-$140&.069~(39)   & & $-$205&.453~(122)       & $-$210&.826~(96)       \\
$\Delta B$                  &     27&.53028~(5)     &     27&.53255~(33)  & &     54&.05732~(11)  &     54&.07405~(54) & &     54&.50273~(7)       &     54&.50740~(90)     \\
$\Delta D_K \times 10^3$    &  $-$11&.46~(48)       &  $-$18&.11~(81)     & &  $-$20&.2~(3)       &  $-$12&.3~(10)     & &   $-$7&.5~(16)          &     20&.1~(42)         \\
$\Delta D_{JK} \times 10^3$ &      0&.9875~(6)      &      0&.8739~(129)  & &      1&.6755~(25)   &      1&.3316~(69)  & &      1&.809~(2)         &      1&.790~(39)       \\
$\Delta D_J \times 10^6$    &     95&.599~(17)      &     93&.436~(69)    & &    216&.32~(4)      &    218&.41~(18)    & &    189&.16~(3)          &    182&.93~(24)        \\
$\Delta H_K \times 10^6$    &     14&.9~(22)        & $-$215&.0~(51)      & &       &             & $-$384&.3~(72)     & &       &                 &       &                \\
$\Delta H_{KJ} \times 10^6$ &      0&.034~(2)       &      0&.908~(114)   & &      0&.150~(24)    &       &            & &      0&.025~(6)         &       &                \\
$\Delta H_{JK} \times 10^9$ &      2&.59~(6)        &     35&.74~(171)    & &     17&.71~(34)     &       &            & &   $-$0&.37~(21)         &       &                \\
$\Delta H_J \times 10^{12}$ &    315&.3~(30)        &       &             & &    200&.7~(63)      &       &            & &    627&.8~(59)          &       &                \\
$\Delta L_J \times 10^{15}$ &   $-$2&.64~(20)$^d$   &       &             & &   $-$5&.28~(40)$^d$ &       &            & &   $-$5&.28~(40)$^d$     &       &                \\
$A\zeta \times 10^{-3}$     &    138&.65620~(7)     &    138&.64715~(7)   & &       &             &       &            & &    138&.65604~(10)      &    138&.63733~(16)     \\
$\eta_K$                    &     10&.333~(7)       &     10&.448~(93)    & &       &             &       &            & &     10&.405~(10)        &     10&.448~(9)        \\
$\eta_J$                    &      0&.390469~(7)    &      0&.389679~(66) & &       &             &       &            & &      0&.39451~(1)       &      0&.39420~(18)     \\
$\eta_{KK} \times 10^6$     & $-$834&.~(41)$^d$     & $-$453&.~(174)      & &       &             &       &            & & $-$834&.~(41)$^d$       &       &                \\
$\eta_{JK} \times 10^6$     &  $-$34&.06~(6)$^d$    &  $-$40&.77~(90)     & &       &             &       &            & &  $-$34&.06~(6)$^d$      &  $-$46&.8~(30)         \\
$\eta_{JJ} \times 10^6$     &   $-$2&.3595~(24)$^d$ &   $-$2&.6750~(111)  & &       &             &       &            & &   $-$2&.3595~(24)$^d$   &   $-$2&.845~(45)       \\
$\eta_{KKK} \times 10^6$    &       &               &     15&.23~(90)     & &       &             &       &            & &       &                 &       &                \\
$\eta_{JKK} \times 10^9$    &      2&.59~(17)$^d$   &       &             & &       &             &       &            & &      2&.59~(17)$^d$     &       &                \\
$\eta_{JJK} \times 10^9$    &      0&.509~(6)$^d$   &       &             & &       &             &       &            & &      0&.509~(6)$^d$     &       &                \\
$q$                         &     17&.79844~(2)     &     17&.79776~(11)  & &       &$^e$         &       &$^e$        & &     17&.72986~(14)$^e$  &     17&.70068~(51)$^e$ \\
$q_{K} \times 10^3$         &   $-$2&.6645~(111)    &       &             & &       &$^e$         &       &$^e$        & &   $-$2&.6153~(89)$^{e}$ &       &                \\
$q_{J} \times 10^6$         &  $-$63&.84~(1)        &  $-$63&.80~(16)     & &       &$^e$         &       &$^e$        & &  $-$68&.67~(3)$^e$      &  $-$66&.84~(17)$^e$    \\
$q_{JK} \times 10^9$        &     93&.19~(53)$^d$   &       &             & &       &$^e$         &       &$^e$        & &     93&.19~(53)$^{d,e}$ &       &                \\
$q_{JJ} \times 10^{12}$     &    312&.~(2)          &    324&.~(57)       & &       &$^e$         &       &$^e$        & &    192&.~(3)$^e$        &       &                \\
\hline \hline
\end{tabular}\\[2pt]
}
\end{center}
{\footnotesize
$^a$ Numbers in parentheses are one standard deviation in units of the least significant figures. 
     Empty entries indicate parameter not applicable or not used in the fit. See section~\ref{det_spec_parameters} for sign and value considerations.\\
$^b$ Parameter difference $\Delta (X) = X_i - X_0$ given for rotational and distortion parameters ($i$ represents an excited vibrational state). 
     Signs were adjusted to present definition for data from Refs.~\cite{MeCN_nu8_1992,MeCN_2nu8_1993}. In all instances, previous values refer 
     to the combined fits of IR and rotational data.\\
$^c$ All parameters given in units of megahertz, except for $E_{\rm{vib}}$, which is given in 
     units of inverse centimeters.\\
$^d$ Ratio kept fixed in the fit for respective $X$ or $\Delta X$.\\
$^e$ The parameter $q$ and its distortion corrections connect levels with $\Delta K = \Delta l = 2$, 
     see subsection~\ref{intro-spec} and section~\ref{det_spec_parameters}.\\
}
\end{table*}

%%%%%%%%%%%%%%%%%%%%%%%%%%%%%%%%%%%%%%%%%%%%%%%%%%%%%%%%%%%%%%%%%%%%%%%%%%%%%%%%%%%%%
%%%%%%%%%%%%%%%%%%%%%%%%%%%%%%%%%%%%%%%%%%%%%%%%%%%%%%%%%%%%%%%%%%%%%%%%%%%%%%%%%%%%%

%%%%%%%%%%%%%%%%%%%%%%%%%%%%%%%%%%%%%%%%%%%%%%%%%%%%%%%%%%%%%%%%%%%%%%%%%%%%%%%%%%%%%
%%%%%  Discussion  %%%%%%%%%%%%%%%%%%%%%%%%%%%%%%%%%%%%%%%%%%%%%%%%%%%%%%%%%%%%%%%%%%
%%%%%%%%%%%%%%%%%%%%%%%%%%%%%%%%%%%%%%%%%%%%%%%%%%%%%%%%%%%%%%%%%%%%%%%%%%%%%%%%%%%%%

\section{Discussion}
\label{Discussion}

Our present ground state spectroscopic parameters are very close to those of our 
previous study \cite{MeCN_rot_2009}; differences outside three times the experimental 
uncertainties occur only in $D_J$, $H_J$, and $L_J$. The experimental data associated 
with the $\varv = 0$/$\varv _8 = 1$ perturbation help to constrain the $K$ level structure 
in the ground vibrational state, additional contributions may come from the data involving 
the $\varv _8 = 1$/$\varv _8 = 2$ perturbations. As a consequence, we obtained a precisely 
determined value for $H_K$ as $(164.6 \pm 6.6)$~Hz. The agreement with $(156 \pm 72)$~Hz 
from an earlier study \cite{MeCN_DeltaK=3_1993} is good, though the large uncertainty in 
the earlier study limits the value of the agreement somewhat. Our previous estimate of 
51~Hz, derived by scaling $D_K$ with the $D_K/A$ ratio as an order-of-magnitude estimate, 
was apparently too small.

As can be seen in \textbf{Table~\ref{comparison_v8_le_2_new-old}}, the extensive new rotational 
data pertaining to $\varv _8 = 1$ and 2 allowed us to determine spectroscopic parameters or 
changes thereof up to higher order than previously \cite{MeCN_nu8_1992,MeCN_2nu8_1993}. 
The uncertainties of many lower order parameters were improved in addition.

The previous and present spectroscopic parameters agree reasonably at lower order and 
particularly for those with dependence on $J$. The agreement deteriorates for higher order 
parameters, especially those dependent on $K$. Consideration of the resonances between 
$\varv _8 = 1$ and 2 as well as between $\varv _8 = 2$ and 3 affected the purely 
$K$-dependent parameters significantly, very similar to the case of propyne, where the 
Fermi interactions involving the equivalent $\nu _{10}$ band occur at slightly lower $K$ 
\cite{MeCCH_nu10+Dyade_2002,MeCCH_nu10+Dyade_2004}. As in the case of propyne, we did not 
require $\eta _{KKK}$ to be included in the fit. Comparison of the various $\eta$ parameters 
from our study as well as those from Ref.~\cite{MeCN_nu8_1992} suggests that the magnitude 
of $\eta _{KKK}$ reported in that study is at least a factor of 100 too large. 
Similarly, the previous $\Delta H_K$ value of $\varv _8 = 1$ in CH$_3$CN was $-215 \pm 5$~Hz 
\cite{MeCN_nu8_1992}, whereas our value is $14.9 \pm 2.2$~Hz; in addition, a value of 
$-384 \pm 7$~Hz was reported for $\varv _8 = 2^0$, in Ref.~\cite{MeCN_nu8_1992}, 
whereas we obtained a value similar to that in $\varv _8 = 1$ in trial fits. 
However, the uncertainties were only slightly smaller than the values, so 
$\Delta H_K(\varv _8 = 2^0)$ was omitted in the final fit.

The vibrational changes in distortion parameters from $\varv = 0$ to $\varv _8 = 1$ in 
\textbf{Table~\ref{parameter_v8_le_2}} are often small, though not in all instances. 
The rather large changes in $H_J$ and $L_J$ are common in these types of molecules. 
In fact, the changes of $H_J$ upon excitation of each of the two low-lying bending modes 
are even larger in magnitude in propyne \cite{MeCCH_nu10+Dyade_2004}. Excitation of the 
CCN bending mode leads also to large changes in $H$ of HC$_3$N \cite{HC3N_rot-v_2000}, 
DC$_3$N \cite{DC3N_rot_2008}, or other, similar molecules.

The changes of rotational and centrifugal distortion parameters from ground state values 
in $\varv _8 = 2^0$ and $\varv _8 = 2^2$ are roughly twice those in $\varv _8 = 1$, 
though pronounced deviations occur already in $\Delta A$ (or $\Delta (A-B)$). The similarity 
to related values in propyne \cite{MeCCH_nu10+Dyade_2002,MeCCH_nu10+Dyade_2004} and methyl 
isocyanide \cite{MeNC_IR_1995,MeNC_IR_rot_2011} may indicate that the low energies of the 
vibrations lead to substantial changes in $\Delta A$ with different $l$ quantum number. 
Changes upon excitation in $\varv _8$ are closer to the ratio of 2 in $\Delta B$, 
but deviations increase, in particular for the changes in the $\Delta H$ values. 
These deviations may be caused by the incomplete analysis of the $\varv _8 = 2$/3 or 
other resonances.

The $A\zeta$ values are essentially identical between $\varv _8 = 1$ and $\varv _8 = 2^2$. 
The changes in $\eta _K$ and $\eta _J$ are rather small, and higher order parameters were 
constrained to the same values. The $q$ values are rather similar also, whereas the $q_{JJ}$ 
values differ considerably, possibly again a consequence of a not or not fully accounted 
interaction.

With respect to the perturbation of the ground vibrational state by $\varv _8 = 1$, 
\v{S}ime\v{c}kov{\'a} et al. \cite{MeCN_rot_2004} calculated perturbations of about 
45~MHz and 36~MHz for the two ground state transitions involving $J = 43$ and $K = 14$, 
this is about 20~MHz larger than actually observed. The difference can be traced almost 
entirely to the energy differences between the two perturbing levels at $J = 43$, which 
was estimated to be 0.013~cm$^{-1}$ \cite{MeCN_rot_2004} whereas our calculations yield 
0.025~cm$^{-1}$. Our value of $F_{ac}(8^{2,\pm2},4^1)/2$, $4.368 \pm 0.001$~MHz 
agrees quite well with $|w_{488}| =4.101 \pm 0.042$~MHz from the combined analysis of 
IR and rotational data or with $4.170 \pm 0.042$~MHz from an analysis of IR data only 
\cite{MeCN_nu4_nu7_3nu8_1993}. We should point out that the sign cannot be determined 
from the fit of the line frequencies, but it may be determinable from relative intensities 
in the rotational or IR spectra. The value given in Ref~\cite{MeCN_nu4_nu7_3nu8_1993} 
was actually negative; we will try to determine the sign in our subsequent study which 
will include $\varv _4 = 1$ data. Other interaction parameters between different 
vibrational states were either determined for the first time here or were not released 
in our fits.

Selected low order spectroscopic parameters of CH$_3$CN have been compared in 
\textbf{Table~\ref{comp_MeX}} with corresponding values of CH$_3$CCH and CH$_3$NC. 
The data of CH$_3$CN  and of CH$_3$CCH are comparatively similar, while those 
of CH$_3$NC with its much lower bending mode are substantially different. The Fermi 
parameters between the first and the second bending state reflect these changes, 
and the values appear to scale approximately with $q_{\rm b}$.

$F(8^{2,\pm2},8^{3,\mp1})$ and $F(8^{2,0},8^{3,3})$ should ideally differ from 
$F(8^{\pm1},8^{2,\mp2})$ by factors of $\sqrt2$ and $\sqrt3$, respectively, see, e.g., 
Ref.~\cite{MeCCH_nu10+Dyade_2004}. The experimental values of $77208 \pm 93$~MHz and 
$91509 \pm 131$~MHz compare quite well with the expected values of 75192~MHz and 
92091~MHz, respectively, in particular considering that there are no $\varv _8 = 3$ 
transition frequencies in the fit, and that we used spectroscopic parameters from a 
previous study for this state \cite{MeCN_nu4_nu7_3nu8_1993}.

The lack of assignments in $2\nu_8 - \nu_8$ may limit the accuracies of the predicted 
line positions in the $2\nu_8^{+2} - \nu_8^{+1}$ sub-band at higher $K$. 
However, our data on the perturbations between $\varv _8 = 1$ and 2 may in part 
compensate these limitations.

%%%%%%%%%%%%%%%%%%%%%%%%%%%%%%%%%%%%%%%%%%%%%%%%%%%%%%%%%%%%%%%%%%%%%%%%%%%%%%%%%%%%%
%%%%%   Table 7  %%%%%%%%%%%%%%%%%%%%%%%%%%%%%%%%%%%%%%%%%%%%%%%%%%%%%%%%%%%%%%%%%%%%
%%%%%%%%%%%%%%%%%%%%%%%%%%%%%%%%%%%%%%%%%%%%%%%%%%%%%%%%%%%%%%%%%%%%%%%%%%%%%%%%%%%%%

\begin{table}
\begin{center}
\caption{Comparison of selected low order spectroscopic parameters or differences $\Delta$ 
         thereof$^{a}$ (cm$^{-1}$, MHz)$^b$ of methyl cyanide, methyl acetylene, and 
         methyl isocyanide in their ground and lowest bending states b.}
\label{comp_MeX}
% \smallskip
{\footnotesize
\begin{tabular}[t]{lr@{}lr@{}lr@{}l}
\hline 
Parameter $X$ & \multicolumn{2}{c}{CH$_3$CN$^c$} & \multicolumn{2}{c}{CH$_3$CCH$^d$} & \multicolumn{2}{c}{CH$_3$NC$^e$} \\
\hline
$A_0$                        & 158\,099&.    & 159\,139&.    & 157\,308&.    \\
$B_0$                        &   9\,198&.9   &   8\,545&.9   &  10\,052&.8   \\
$D_{K,0} \times 10^3$        &   2\,831&.    &   2\,939&.    &   2\,567&.    \\
$D_{JK,0} \times 10^3$       &      177&.4   &      163&.4   &      227&.5   \\
$D_{J,0} \times 10^6$        &   3\,808&.    &   2\,908&.    &   4\,692&.    \\
$E_{\rm{b}}$$^b$             &      365&.0   &      330&.9   &      267&.3   \\
$\Delta ^a A_{\rm b}$        &    $-$88&.3   &    $-$61&.9   &        0&.6   \\
$\Delta ^a B_{\rm b}$        &       27&.53  &       23&.94  &       39&.07  \\
$A\zeta_{\rm b}$             & 138\,656&.    & 141\,919&.    & 146\,007&.    \\
$\eta_{K,\rm b}$             &       10&.34  &       10&.94  &       10&.13  \\
$\eta_{J,\rm b} \times 10^3$ &      390&.5   &      334&.8   &      587&.6   \\
$q_{\rm b}$                  &       17&.80  &       16&.79  &        6&.94  \\
$F$(b$^{\pm1}$,b$^{2,\mp2}$) &  53\,169&.    &  51\,745&.    &  23\,808&.    \\
\hline \hline
\end{tabular}\\[2pt]
}
\end{center}
{\footnotesize
$^a$ Parameter $X$ given for $\varv = 0$; $\Delta (X) = X_i - X_0$, with $i$ representing 
     an excited vibrational state.\\
$^b$ All parameters given in units of megahertz, except for $E_{\rm{b}}$, which is given in 
     units of inverse centimeters.\\
$^c$ This work.\\
$^d$ Ref.~\cite{MeCCH_nu10+Dyade_2004}; sign of $q$ adjusted, see section~\ref{det_spec_parameters}.\\
$^e$ Refs.~\cite{CH3NC_v8=1_2_2011,MeNC_IR_1995,MeNC_IR_rot_2011}; sign of $q$ adjusted, 
     see section~\ref{det_spec_parameters}.\\
}
\end{table}

%%%%%%%%%%%%%%%%%%%%%%%%%%%%%%%%%%%%%%%%%%%%%%%%%%%%%%%%%%%%%%%%%%%%%%%%%%%%%%%%%%%%%
%%%%%%%%%%%%%%%%%%%%%%%%%%%%%%%%%%%%%%%%%%%%%%%%%%%%%%%%%%%%%%%%%%%%%%%%%%%%%%%%%%%%%

%%%%%%%%%%%%%%%%%%%%%%%%%%%%%%%%%%%%%%%%%%%%%%%%%%%%%%%%%%%%%%%%%%%%%%%%%%%%%%%%%%%%%
%%%%%  Database Issues  %%%%%%%%%%%%%%%%%%%%%%%%%%%%%%%%%%%%%%%%%%%%%%%%%%%%%%%%%%%%%
%%%%%%%%%%%%%%%%%%%%%%%%%%%%%%%%%%%%%%%%%%%%%%%%%%%%%%%%%%%%%%%%%%%%%%%%%%%%%%%%%%%%%

\section{Expanded database for CH$_3$CN line parameters}
\label{database_issues}

Compilations of CH$_3$CN molecular line parameters are greatly expanded by including new or 
updated predictions of spectra within or between the three states of the present investigation. 
Predictions of rotational and infrared spectra for astrophysical applications will be available 
in the catalog section\footnote{https://cdms.astro.uni-koeln.de/classic/entries/} of the 
CDMS~\cite{CDMS_1,CDMS_2} and that of JPL\footnote{http://spec.jpl.nasa.gov/ftp/pub/catalog/catdir.html}
\cite{JPL-catalog_1998}. The parameter, line, fit and other auxiliary files are available 
online as supplements with the publication and in the Cologne Spectroscopy Data 
section\footnote{https://cdms.astro.uni-koeln.de/classic/predictions/daten/CH3CN/CH3CN/} 
of the CDMS. For remote sensing of, e.g., Titan, the new compilation will also be provided 
to HITRAN \cite{HITRAN_2012} and GEISA \cite{GEISA_2009} with N$_2$-pressure and 
self-broadening information included. For this, the extensive pressure broadening CH$_3$CN 
measurements of $P$ and $R$ branch transitions of $\nu_4$ \cite{MeCN_nu4_int_2008} 
were modeled by extending \textit{empirical expressions} applied to ammonia 
\cite{NH3_2mu_1996,NH3_10mu_2004}. The resulting polynomial has no physical meaning, 
but it does provide reasonable estimates of Lorentz broadening coefficients for CH$_3$CN 
transitions.

%%%%%%%%%%%%%%%%%%%%%%%%%%%%%%%%%%%%%%%%%%%%%%%%%%%%%%%%%%%%%%%%%%%%%%%%%%%%%%%%%%%%%
%%%%%  Pressure broadening  %%%%%%%%%%%%%%%%%%%%%%%%%%%%%%%%%%%%%%%%%%%%%%%%%%%%%%%%%
%%%%%%%%%%%%%%%%%%%%%%%%%%%%%%%%%%%%%%%%%%%%%%%%%%%%%%%%%%%%%%%%%%%%%%%%%%%%%%%%%%%%%

%%%%%%%%%%%%%%%%%%%%%%%%%%%%%%%%%%%%%%%%%%%%%%%%%%%%%%%%%%%%%%%%%%%%%%%%%%%%%%%%%%%%%
%%%%%   Table 8  %%%%%%%%%%%%%%%%%%%%%%%%%%%%%%%%%%%%%%%%%%%%%%%%%%%%%%%%%%%%%%%%%%%%
%%%%%%%%%%%%%%%%%%%%%%%%%%%%%%%%%%%%%%%%%%%%%%%%%%%%%%%%%%%%%%%%%%%%%%%%%%%%%%%%%%%%%

\begin{table}
\begin{center}
\caption{Examples of measured$^a$ (obs.; with uncertainties unc. in percent in parentheses) 
         and modeled (calc.; with percent difference in parentheses) $\nu _4$ self-broadened 
         line widths (cm$^{-1}$/atm) of methyl cyanide at 296~K with frequencies (cm$^{-1}$) 
         and quantum numbers.}
\label{width-modeling}
% \smallskip
{\footnotesize
\begin{tabular}[t]{rrcccr}
\hline 
$J'$, $K'$ & $J''$, $K''$ & Frequency & \multicolumn{2}{c}{Self-broadened width} & $J_m$, $K_m$ \\
      &       &           & obs.~(\%unc.) & calc.~(\%diff.) &       \\
\hline
 2, 1 &  3, 1 & 918.43291 & 1.358~(0.7) & 1.2588~~~(7.88)  &  3, 1 \\
 3, 1 &  2, 1 & 922.10564 & 1.307~(1.0) & 1.2588~~~(3.83)  &  3, 1 \\
 7, 6 &  8, 6 & 915.09499 & 1.071~(0.7) & 1.0798~($-$0.82) &  8, 6 \\
 8, 6 &  7, 6 & 924.88187 & 1.039~(0.9) & 1.0798~($-$3.78) &  8, 6 \\
18, 9 & 19, 9 & 907.67687 & 1.722~(0.5) & 1.6957~~~(1.55)  & 19, 9 \\
19, 9 & 18, 9 & 930.89430 & 1.711~(0.7) & 1.6957~~~(0.90)  & 19, 9 \\
22, 7 & 23, 7 & 905.12404 & 1.773~(0.4) & 1.7497~~~(1.33)  & 23, 7 \\
23, 7 & 22, 7 & 933.24349 & 1.821~(0.6) & 1.7497~~~(4.07)  & 23, 7 \\
28, 0 & 29, 0 & 901.25148 & 1.583~(0.5) & 1.5807~~~(0.15)  & 29, 0 \\
29, 0 & 28, 0 & 936.73087 & 1.528~(0.5) & 1.5807~($-$3.33) & 29, 0 \\
40, 4 & 41, 4 & 892.54529 & 0.896~(0.7) & 0.8737~~~(2.55)  & 41, 4 \\
41, 4 & 40, 4 & 942.65553 & 0.823~(0.8) & 0.8737~($-$5.81) & 41, 4 \\
\hline \hline
\end{tabular}\\[2pt]
}
\end{center}
{\footnotesize
$^a$ The measured values are taken from Ref.~\cite{MeCN_nu4_int_2008}.\\
}
\end{table}

%%%%%%%%%%%%%%%%%%%%%%%%%%%%%%%%%%%%%%%%%%%%%%%%%%%%%%%%%%%%%%%%%%%%%%%%%%%%%%%%%%%%%
%%%%%%%%%%%%%%%%%%%%%%%%%%%%%%%%%%%%%%%%%%%%%%%%%%%%%%%%%%%%%%%%%%%%%%%%%%%%%%%%%%%%%
\subsection{Pressure-broadened half widths}
\label{Pressure-broadening}
%%%%%%%%%%%%%%%%%%%%%%%%%%%%%%%%%%%%%%%%%%%%%%%%%%%%%%%%%%%%%%%%%%%%%%%%%%%%%%%%%%%%%

The measured N$_2$- and self-broadened halfwidths (HWHM) for CH$_3$CN in the $\nu_4$ 
band at 920~cm$^{-1}$ vary smoothly with respect to their quantum numbers $J$ and $K$ 
\cite{MeCN_nu4_int_2008}. Patterns become more obvious when the widths are examined 
as a function of values related to the transition quantum numbers; these are defined 
as $J_m = J'' + 1$, $K_m = K'' + 1$ for $R$ branch transitions, and $J_m = J''$, 
$K_m = K''$ for $P$ and $Q$ branch transitions. Examples are shown in 
\textbf{Table~\ref{width-modeling}}. N$_2$- and self-broadened half widths were derived 
for our CH$_3$CN compilation by modeling the reported widths for the $\nu_4$ band 
\cite{MeCN_nu4_int_2008} according to the expression in Eq.~\ref{polynomial}; 
the polynomial coefficients are listed in \textbf{Table~\ref{coeff-Tab}}.

\begin{equation}
\label{polynomial}
\gamma(J_m, K_m) = \sum_{i, j} a_{ij}J_m^iK_m^j
\end{equation}

%%%%%%%%%%%%%%%%%%%%%%%%%%%%%%%%%%%%%%%%%%%%%%%%%%%%%%%%%%%%%%%%%%%%%%%%%%%%%%%%%%%%%
%%%%%   Table 9  %%%%%%%%%%%%%%%%%%%%%%%%%%%%%%%%%%%%%%%%%%%%%%%%%%%%%%%%%%%%%%%%%%%%
%%%%%%%%%%%%%%%%%%%%%%%%%%%%%%%%%%%%%%%%%%%%%%%%%%%%%%%%%%%%%%%%%%%%%%%%%%%%%%%%%%%%%

\begin{table}
\begin{center}
\caption{Coefficients to model the N$_2$- and self-broadened widths of the $\nu _4$ band  
         of CH$_3$CN using Eq.~\ref{polynomial}.}
\label{coeff-Tab}
% \smallskip
{\footnotesize
\begin{tabular}[t]{lr@{}lr@{}l}
\hline 
Coefficients & \multicolumn{2}{c}{N$_2$} & \multicolumn{2}{c}{Self} \\
\hline
$a_{00}$             &     0&.171    &      1&.621    \\
$a_{10} \times 10^3$ &     4&.731    & $-$172&.4      \\
$a_{01} \times 10^3$ &  $-$9&.962    &  $-$82&.99     \\
$a_{20} \times 10^3$ &  $-$0&.7152   &     26&.32     \\
$a_{11} \times 10^3$ &     1&.704    &     18&.89     \\
$a_{02} \times 10^3$ &  $-$1&.233    &  $-$14&.42     \\
$a_{30} \times 10^3$ &     0&.03155  &   $-$1&.28     \\
$a_{21} \times 10^3$ &  $-$0&.1068   &   $-$1&.296    \\
$a_{12} \times 10^3$ &     0&.1361   &      0&.7237   \\
$a_{03} \times 10^3$ &     0&.002544 &      0&.1243   \\
$a_{40} \times 10^6$ &  $-$0&.6094   &     24&.83     \\
$a_{31} \times 10^6$ &     2&.643    &  $-$34&.66     \\
$a_{22} \times 10^6$ &  $-$3&.849    &  $-$10&.5      \\
$a_{13} \times 10^6$ &  $-$1&.819    &      9&.179    \\
$a_{50} \times 10^9$ &     4&.388    & $-$171&.6      \\
$a_{41} \times 10^9$ & $-$22&.42     & $-$319&.8      \\
$a_{32} \times 10^9$ &    33&.41     &     66&.99     \\
$a_{23} \times 10^9$ &    37&.85     & $-$430&.9      \\
\hline \hline
\end{tabular}
}
\end{center}
\end{table}

%%%%%%%%%%%%%%%%%%%%%%%%%%%%%%%%%%%%%%%%%%%%%%%%%%%%%%%%%%%%%%%%%%%%%%%%%%%%%%%%%%%%%
%%%%%%%%%%%%%%%%%%%%%%%%%%%%%%%%%%%%%%%%%%%%%%%%%%%%%%%%%%%%%%%%%%%%%%%%%%%%%%%%%%%%%

\begin{figure*}
 \begin{center}
  \includegraphics[width=16cm]{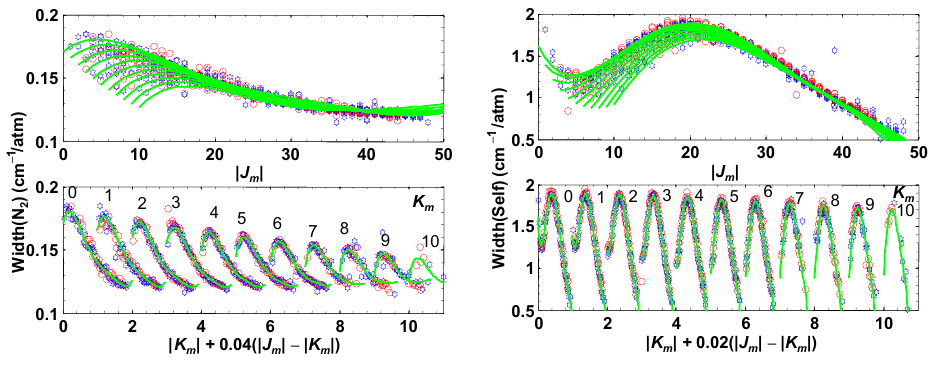}
 \end{center}
  \caption{Model fit of measured N$_2$- (left) and self-broadening (right) Lorentz width 
           coefficients in cm$^{-1}$/atm at 296~K for the $\nu _4$ band \cite{MeCN_nu4_int_2008}. 
           Many of the widths are well reproduced by Eq.~\ref{polynomial}. The solid lines 
           in green represent calculated values while observed widths of $P$ and $R$ branch 
           transitions are indicated by circles in red and hexagons in blue, respectively. 
           In the lower parts, the $J_m$ and $K_m$ values are combined for the horizontal 
           axes to reveal the variation patterns.}
  \label{fig_line-width_N2_self}
 \end{figure*}

The measured widths were weighted in the fit by the inverse of the square of the reported 
uncertainties \cite{MeCN_nu4_int_2008}. The impact of a few outliers (at very low and 
higher $J_m$ and $K_m$) were suppressed as an important side-effect. As shown in 
\textbf{Fig.~\ref{fig_line-width_N2_self}}, the calculated values using Eq.~\ref{polynomial} 
(solid curves) reproduce the pattern of the measured data points within 5\,\% for most 
of the transitions. The full lists are given in Supplemental files 
5\_Width(N2)\_CH3CN\_v4.out and 6\_Width(Self)\_CH3CN\_v4.out. 
Finally, we adopted this expression to estimate N$_2$- and self-broadened half widths 
for the three IR bands, $\nu_8$, $2\nu_8 - \nu_8$, $2\nu_8$ of the present study. 
Because the experimental $\nu _4$ lines, for which line widths were determined, 
have $J_m \lesssim 50$, we limited the line-widths at 296~K to:

$0.12 < \gamma_{\rm{N}_2} < 0.2$~cm$^{-1}$/atm for N$_2$-broadened half-widths, and

$0.2 < \gamma_{\rm{self}} < 2.0$~cm$^{-1}$/atm for self-broadened half-widths.

%%%%%%%%%%%%%%%%%%%%%%%%%%%%%%%%%%%%%%%%%%%%%%%%%%%%%%%%%%%%%%%%%%%%%%%%%%%%%%%%%%%%%
%%%%%  Figure 10  %%%%%%%%%%%%%%%%%%%%%%%%%%%%%%%%%%%%%%%%%%%%%%%%%%%%%%%%%%%%%%%%%%%
%%%%%%%%%%%%%%%%%%%%%%%%%%%%%%%%%%%%%%%%%%%%%%%%%%%%%%%%%%%%%%%%%%%%%%%%%%%%%%%%%%%%%

%%%%%%%%%%%%%%%%%%%%%%%%%%%%%%%%%%%%%%%%%%%%%%%%%%%%%%%%%%%%%%%%%%%%%%%%%%%%%%%%%%%%%
%%%%%%%%%%%%%%%%%%%%%%%%%%%%%%%%%%%%%%%%%%%%%%%%%%%%%%%%%%%%%%%%%%%%%%%%%%%%%%%%%%%%%

We adopted one single value, 0.72, from the HITRAN2012 database \cite{HITRAN_2012} 
for the temperature dependence exponent of the N$_2$-broadened Lorentz widths. 
This value is based on the measurements of the $J = 12 - 11$, $K = 3$ transition 
in the ground vibrational state \cite{MeCN_broadening-v0_2006}.

%%%%%%%%%%%%%%%%%%%%%%%%%%%%%%%%%%%%%%%%%%%%%%%%%%%%%%%%%%%%%%%%%%%%%%%%%%%%%%%%%%%%%
%%%%%  Pressure shifts  %%%%%%%%%%%%%%%%%%%%%%%%%%%%%%%%%%%%%%%%%%%%%%%%%%%%%%%%%%%%%
%%%%%%%%%%%%%%%%%%%%%%%%%%%%%%%%%%%%%%%%%%%%%%%%%%%%%%%%%%%%%%%%%%%%%%%%%%%%%%%%%%%%%
\subsection{N$_2$-pressure-induced frequency shifts}
\label{Pressure-shifts}

Rinsland et al. \cite{MeCN_nu4_int_2008} also reported N$_2$-pressure-induced frequency 
shifts for CH$_3$CN in its $\nu _4$ band, but no distinctive pattern is seen for pressure 
shifts with respect to $J_m$ and $K_m$. However, there was a significant correlation 
between measured line widths and frequency shifts, as presented in their Fig.~11 of 
Ref~\cite{MeCN_nu4_int_2008}. We use this observation in our present work to obtain 
frequency shift estimates for the transition of $\nu_8$, $2\nu_8 - \nu_8$, and $2\nu_8$. 
First, we apply a scaled set of the \textit{measured} N$_2$-induced frequency shifts 
for $\nu _4$ transitions for all other CH$_3$CN transitions having the same set of 
quantum numbers, $J''$, $K''$, $J'$, and $K'$. However, not all transitions predicted 
in the present work have counter parts among the measured $\nu _4$ transitions; we assume 
for these cases that their N$_2$-induced frequency shifts $\delta_4(J_m, K_m)$ can be 
estimated from their pressure-broadened line widths using the \textit{width-shift 
relationship} proposed by Rinsland et al. \cite{MeCN_nu4_int_2008}. Based on Fig.~11 of 
Ref.~\cite{MeCN_nu4_int_2008}, a simple expression was derived as

\begin{equation}
\label{freq-shift}
\delta_4(J_m, K_m) = [\gamma(J_m, K_m)_{{\rm N}_2} - 0.15] / 15 
\end{equation}

One commonly assumes that the overall magnitudes of the pressure shifts increase with 
increasing vibrational band centers \cite{pressure-broadening-theory_1958}. We estimated 
the pressure shifts $\delta_i(J_m, K_m)$ of $\nu _i$ from those of the $\nu _4$ band 
according to

\begin{equation}
\label{freq-shift_other}
\delta_i(J_m, K_m) = \delta_4(J_m, K_m) \times (E_{\rm vib}(i) / E_{\rm vib}(4))
\end{equation}

with $E_{\rm vib}(i)$ being the vibrational energy of $\nu_8$, $2\nu_8 - \nu_8$, or $2\nu_8$ 
and $E_{\rm vib}(4)$ that of $\nu _4$.

\begin{figure*}
 \begin{center}
  \includegraphics[width=16cm]{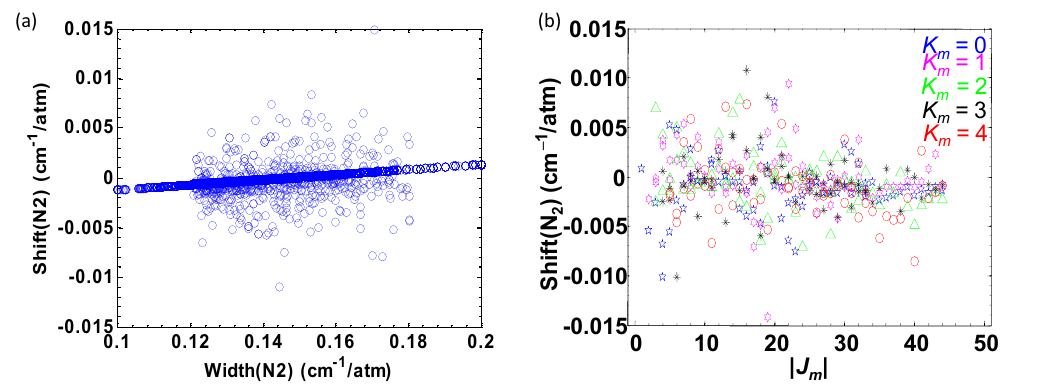}
 \end{center}
  \caption{(a) N$_2$-pressure induced frequency shifts derived for the $\nu _8$ region. 
           The scattered values are adapted from the $\nu _4$ measurements \cite{MeCN_nu4_int_2008} 
           while the well-aligned ones are based on Eqs.~\ref{freq-shift} and \ref {freq-shift_other}. 
           (b) The derived N$_2$-shifts of IR bands in the present study are presented for selected 
           $K_m$ (see text for details).}
  \label{fig_shift-issues}
 \end{figure*}

The resulting N$_2$-pressure shifts for the $\nu_8$, $2\nu_8 - \nu_8$, and $2\nu_8$ bands 
are presented in \textbf{Fig.~\ref{fig_shift-issues}a}. The width-shift relationships resemble 
those observed in the $\nu _4$ band \cite{MeCN_nu4_int_2008}. The distribution with respect to 
$K_m$, shown in \textbf{Fig.~\ref{fig_shift-issues}b}, does not show a distinctive pattern, 
as was the case for $\nu _4$ \cite{MeCN_nu4_int_2008}.

%%%%%%%%%%%%%%%%%%%%%%%%%%%%%%%%%%%%%%%%%%%%%%%%%%%%%%%%%%%%%%%%%%%%%%%%%%%%%%%%%%%%%
%%%%%  Figure 11  %%%%%%%%%%%%%%%%%%%%%%%%%%%%%%%%%%%%%%%%%%%%%%%%%%%%%%%%%%%%%%%%%%%
%%%%%%%%%%%%%%%%%%%%%%%%%%%%%%%%%%%%%%%%%%%%%%%%%%%%%%%%%%%%%%%%%%%%%%%%%%%%%%%%%%%%%

%%%%%%%%%%%%%%%%%%%%%%%%%%%%%%%%%%%%%%%%%%%%%%%%%%%%%%%%%%%%%%%%%%%%%%%%%%%%%%%%%%%%%
%%%%%  Intensities  %%%%%%%%%%%%%%%%%%%%%%%%%%%%%%%%%%%%%%%%%%%%%%%%%%%%%%%%%%%%%%%%%
%%%%%%%%%%%%%%%%%%%%%%%%%%%%%%%%%%%%%%%%%%%%%%%%%%%%%%%%%%%%%%%%%%%%%%%%%%%%%%%%%%%%%
\subsection{Intensities of $\nu_8$, $2\nu_8 - \nu_8$, and $2\nu_8$}
\label{intensity_issues}

We evaluated transition dipole moments for each of the three IR bands of the present study. 
There are some indications that additional correction terms may lead to a better intensity 
modelling, especially in the cases of $\nu_8$ and $2\nu_8 - \nu_8$. More sophisticated 
intensity modelling is beyond the aim of the present study, in particular because the 
quality of the IR spectrum in the $\nu_8$ region is not sufficient for such modelling.

%%%%%%%%%%%%%%%%%%%%%%%%%%%%%%%%%%%%%%%%%%%%%%%%%%%%%%%%%%%%%%%%%%%%%%%%%%%%%%%%%%%%%
%%%%%   Table 10  %%%%%%%%%%%%%%%%%%%%%%%%%%%%%%%%%%%%%%%%%%%%%%%%%%%%%%%%%%%%%%%%%%%
%%%%%%%%%%%%%%%%%%%%%%%%%%%%%%%%%%%%%%%%%%%%%%%%%%%%%%%%%%%%%%%%%%%%%%%%%%%%%%%%%%%%%

\begin{table*}
\begin{center}
\caption{Derived transition dipole moments $\mu$ (D) and integrated intensities 
         (10$^{-19}$~cm/molecule) of IR bands of $^{12}$CH$_3 ^{12}$C$^{14}$N calculated 
         from the present Hamiltonian model with respect to the transition dipole moments 
         (Calc.~($\mu$)) and apparent intensities in the $l$ components (Calc.~($l$)) 
         together with the extrapolated and experimentally measured total integrated 
         intensities from previous studies.}
\label{mu-IR_intensities}
% \smallskip
{\footnotesize
\begin{tabular}[t]{lr@{}lr@{}lr@{}lr@{}lr@{}lr@{}lr@{}l}
\hline
        &  &   \multicolumn{12}{c}{Integrated intensity} \\
\cline{4-15}
IR band & \multicolumn{2}{c}{$\mu$} & \multicolumn{2}{c}{Calc.~($\mu$)} & \multicolumn{2}{c}{Calc.~($l$)} 
& \multicolumn{2}{c}{Ref.~\cite{MeCN-int_2005}} & \multicolumn{2}{c}{Ref.~\cite{MeCN-int_1995}} 
& \multicolumn{2}{c}{Ref.~\cite{MeCN-int_1985}} & \multicolumn{2}{c}{Ref.~\cite{MeCN-int_1984}} \\
\hline
$\varv _8 = 1 - 0$   & 0&.043   & 1&.483     & 1&.483  &  &         &  &        &  &      &  &         \\
$\varv _8 = 2^0 - 1$ & 0&.043   & 0&.251     & 0&.244  &  &         &  &        &  &      &  &         \\
$\varv _8 = 2^2 - 1$ & 0&.043   & 0&.257     & 0&.264  &  &         &  &        &  &      &  &         \\
$\nu _8$ region      &  &       & 2&.285$^a$ &  &      &  &         & 1&.77~(4) & 1&.81   & 2&.79~(28) \\
$\varv _8 = 2^0 - 0$ & 0&.030   & 1&.674     & 1&.585  &  &         &  &        &  &      &  &         \\
$\varv _8 = 2^2 - 0$ & 0&.000   & 0&.000     & 0&.090  &  &         &  &        &  &      &  &         \\
$2\nu _8$ region     &  &       & 2&.580$^a$ &  &      & 2&.63~(18) & 2&.50~(4) & 3&.43   &  &         \\

\hline \hline
\end{tabular}\\[2pt]
}
\end{center}
{\footnotesize
$^a$ Estimated from the cold bands, $\varv _8 = 1 - 0$ and $2 - 0$, by multiplication with the 
     vibrational partition factor 1.5006 at 296~K. An increase of 2.677\,\% from other isotopic species 
     was considered.}
\end{table*}

%%%%%%%%%%%%%%%%%%%%%%%%%%%%%%%%%%%%%%%%%%%%%%%%%%%%%%%%%%%%%%%%%%%%%%%%%%%%%%%%%%%%%
%%%%%%%%%%%%%%%%%%%%%%%%%%%%%%%%%%%%%%%%%%%%%%%%%%%%%%%%%%%%%%%%%%%%%%%%%%%%%%%%%%%%%
\subsubsection{$\nu_8$}
%%%%%%%%%%%%%%%%%%%%%%%%%%%%%%%%%%%%%%%%%%%%%%%%%%%%%%%%%%%%%%%%%%%%%%%%%%%%%%%%%%%%%
\label{nu8-intenity}

Our simulations on a line-by-line basis indicate that the $\nu_8$ spectrum is represented 
reasonably well by a transition dipole moment of $\mu(8^1 _0) = 0.043$~D.

We calculated partition function values by summing over all states up to $J = 99$, $K = 25$, 
and $\varv _8 = 3$. While the rotational contribution to the partition function at 296~K 
is well converged, the vibrational contribution is slightly truncated. We have estimated 
the truncation error classically, assuming a rigid rotor as well as a harmonic oscillator. 
This approximation in itself introduces some error, but we estimate it to be small 
with respect to the truncation error. The truncation leads to a corrections of 1.408\,\%. 
The resulting intensity, integrated over all lines, is given in Table~\ref{mu-IR_intensities}. 
The integrated intensity in the entire $\nu _8$ region, i.e. with all hot bands, was 
estimated by multiplying the calculated $\nu _8$ intensity with the vibrational partition 
factor of 1.5006. An additional increase of the integrated intensity by 2.677\,\% was 
required because of the contributions of other isotopic species. There is a convention 
in HITRAN that intensities refer to a sample in natural isotopic composition. 
The resulting intensity is also given in Table~\ref{mu-IR_intensities}. Our value is about 
half way between two lower values \cite{MeCN-int_1995,MeCN-int_1985} and one higher one 
\cite{MeCN-int_1984}.

%%%%%%%%%%%%%%%%%%%%%%%%%%%%%%%%%%%%%%%%%%%%%%%%%%%%%%%%%%%%%%%%%%%%%%%%%%%%%%%%%%%%%
\subsubsection{$2\nu_8 - \nu_8$}
%%%%%%%%%%%%%%%%%%%%%%%%%%%%%%%%%%%%%%%%%%%%%%%%%%%%%%%%%%%%%%%%%%%%%%%%%%%%%%%%%%%%%

\begin{figure}
 \begin{center}
  \includegraphics[width=8.0cm]{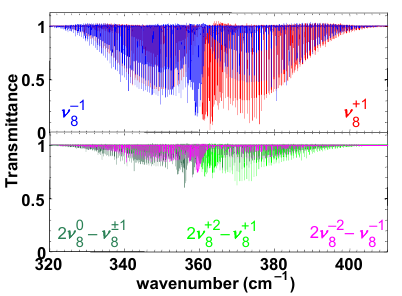}
 \end{center}
  \caption{Contributions of each sub-band in the $\nu _8$ region of CH$_3$CN showing  
           their relative intensities for the cold band in the upper part and for the 
           hot band in the lower part. Note: hot bands originating in states higher 
           than $\varv _8 = 1$ contribute to $\sim$10\,\%  to the total intensity.}
  \label{nu8-modelling}
 \end{figure}

We assumed transition dipole moments of $\mu(8^{2^l} _1) = 0.043$~D for both $l = 0$ and 
$l = \pm2$ components as for the cold band. This approximation appears to be valid well 
within the estimated intensity uncertainties of 4\,\%, see subsubsection~\ref{2nu8-intensity}. 
\textbf{Fig.~\ref{nu8-modelling}} shows the calculated contributions of the $\nu _8$ 
and $2\nu _8 - \nu _8$ subbranches.

As can be seen in Table~\ref{mu-IR_intensities}, slightly more than 10\,\% of the total 
intensity in the $\nu _8$ region of methyl cyanide are caused by IR transitions originating 
in states higher than $\varv _8 = 1$. Finding unblended lines among these will be difficult 
because this was already not easy for $2\nu _8^2 - \nu _8$.

%%%%%%%%%%%%%%%%%%%%%%%%%%%%%%%%%%%%%%%%%%%%%%%%%%%%%%%%%%%%%%%%%%%%%%%%%%%%%%%%%%%%%
\subsubsection{$2\nu_8$}
%%%%%%%%%%%%%%%%%%%%%%%%%%%%%%%%%%%%%%%%%%%%%%%%%%%%%%%%%%%%%%%%%%%%%%%%%%%%%%%%%%%%%
\label{2nu8-intensity}

The relative precisions of measured intensities were checked using spectral comparisons. 
For example, the 731$-$732~cm$^{-1}$ interval in \textbf{Fig.~\ref{2nu8_good_positions}} 
has 181 retrieved features with intensities ranging from $1.3\times10^{-6}$ to 
$7.3\times10^{-3}$~cm$^{-2}$/atm at 293~K. Since the sample was a mixture, line 
intensities of the two calibration species were measured to obtain the partial pressures 
of OCS (3\,\%) and CO$_2$ (0.4\,\%). In addition, selected strong CH$_3$CN lines 
of the $\nu _4$ band were retrieved and compared to values reported in Table~2 of 
Ref.~\cite{MeCN_nu4_int_2008}; this indicated that the retrieved intensities in the 
2$\nu_8$ region should be increased by 4\,\%; with these calibrations, the overall 
(relative) precision is thought to be 0.0002~cm$^{-1}$ for line centers and 5\,\% 
for intensities. However, worse precisions are expected for blended features separated 
by less than 0.0025~cm$^{-1}$ and for $\sim$5100 absorptions weaker than 
$9 \times 10^{-5}$~cm$^{-2}$/atm at 293~K.

Our simulations indicate that the spectrum is represented well by a transition dipole 
moment of $\mu (8^{2^0} _0) = 0.030$~D.  See subsection~\ref{exptl_IR} for considerations 
concerning the experimental intensities and subsubsection~\ref{nu8-intenity} for a 
remark on the partition function and on the sample composition. The integrated 
intensity from our Hamiltonian model is $1.75 \times 10^{-19}$~cm/molecule for the 
$2\nu _8$ band alone. Only a small portion of this intensity, $\sim1.0 \times 10^{-21}$ 
and $\sim8.1 \times 10^{-21}$~cm/molecule is transfered through the $q_{22}$ interaction 
to the $\varv _8 = 2^{+2}$ and $\varv _8 = 2^{-2}$ components, respectively.

About 23\,\% of the $2\nu _8$ intensity at 296~K should arise from the $3\nu _8 - \nu _8$ 
and $\nu _7 - \nu _8$ bands. A preliminary analysis suggests that a fair fraction of 
these lines will not be blended by other lines of similar or higher intensity. 
Even some of the lines originating in states higher than $\varv _8 = 1$, which account 
for slightly more than 10\,\% of the band intensity, may be usable for spectroscopic 
analyses. More work is needed to clarify these aspects.

%%%%%%%%%%%%%%%%%%%%%%%%%%%%%%%%%%%%%%%%%%%%%%%%%%%%%%%%%%%%%%%%%%%%%%%%%%%%%%%%%%%%%
%%%%%  Conclusions  %%%%%%%%%%%%%%%%%%%%%%%%%%%%%%%%%%%%%%%%%%%%%%%%%%%%%%%%%%%%%%%%%
%%%%%%%%%%%%%%%%%%%%%%%%%%%%%%%%%%%%%%%%%%%%%%%%%%%%%%%%%%%%%%%%%%%%%%%%%%%%%%%%%%%%%
\section{Conclusions}
\label{Conclusions}

Our extensive and accurate rotational data in vibrational states $\varv _8 \le 2$ 
resulted in a parameter set extended to higher order and to mostly improved uncertainties 
at lower order. More importantly, these data enabled the first treatments of resonances 
between $\varv _8 = 1$ and 2, as well as between $\varv _8 = 2$ and 3. These resonances 
cause considerable perturbations in the rotational spectrum, and their successful 
treatments affect the spectroscopic parameters, in particular the $K$-dependent ones 
and those at higher order. We have also performed the first in-depth analysis of the 
resonant interaction between the ground vibrational state and $\varv _8 = 1$.

The analysis of a new $2\nu _8$ spectrum was an additional important contribution to 
the determination of spectroscopic parameters. Only about half of the roughly 12000 
lines in the $2\nu _8$ spectrum are caused by the cold band. A large fraction 
of the unaccounted lines will be caused by the $3\nu _8 - \nu _8$ hot band and the 
$\nu_7 - \nu_8$ difference band. Accounting for these will require in depth analysis 
of the $\nu _7$/$3\nu _8$ band system aided by rotational transitions. A still 
considerable fraction of lines are due at least in part to bands originating in 
states higher than $\varv _8 = 1$.

We have derived line broadening and shift parameters for the $\nu_8$, $2\nu_8 - \nu_8$, 
and $2\nu_8$ bands from previously determined $\nu_4$ values and have used this information 
to determine transition moments and intensities for the three IR bands. Individually 
measured transition intensities may lead to an improved intensity model.

We have generated an expanded database for CH$_3$CN line parameters for transitions 
within each of the states $\varv _8 \le 2$, mainly for radio astronomical observations, 
in particular with ALMA and other interferometers. In addition, information is provided 
for transitions between these three states for remote sensing of, e.g., the atmosphere 
of Titan.

The inclusion of data pertaining to $\varv _4 = 1$ is almost complete. We intend to improve 
the data involving its interactions with other vibrational states. We want to study also the 
proposed strong, although distant, thus non-resonant Fermi interaction of $\varv _4 = 1$ with 
$\varv _8 = 2$ \cite{FF_Duncan_1978}. In addition, we have already made assignments 
in the $\nu _4 + \nu _8 - \nu _8$ IR band as well as for rotational transitions in 
$\varv _4 = \varv _8 = 1$, though these are often tentative. The spectra are perturbed, 
but, as it appears, only locally, such that a reasonable deperturbation may be possible 
by considering only spectroscopic parameters of the perturbing states.

%%%%%%%%%%%%%%%%%%%%%%%%%%%%%%%%%%%%%%%%%%%%%%%%%%%%%%%%%%%%%%%%%%%%%%%%%%%%%%%%%%%%%
%%%%%  Figure 12  %%%%%%%%%%%%%%%%%%%%%%%%%%%%%%%%%%%%%%%%%%%%%%%%%%%%%%%%%%%%%%%%%%%
%%%%%%%%%%%%%%%%%%%%%%%%%%%%%%%%%%%%%%%%%%%%%%%%%%%%%%%%%%%%%%%%%%%%%%%%%%%%%%%%%%%%%

%%%%%%%%%%%%%%%%%%%%%%%%%%%%%%%%%%%%%%%%%%%%%%%%%%%%%%%%%%%%%%%%%%%%%%%%%%%%%%%%%%%%%
%%%%%%%%%%%%%%%%%%%%%%%%%%%%%%%%%%%%%%%%%%%%%%%%%%%%%%%%%%%%%%%%%%%%%%%%%%%%%%%%%%%%%


%%%%%%%%%%%%%%%%%%%%%%%%%%%%%%%%%%%%%%%%%%%%%%%%%%%%%%%%%%%%%%%%%%%%%%%%%%%%%%%%%%%%%
%%%%%  acknowledgements  %%%%%%%%%%%%%%%%%%%%%%%%%%%%%%%%%%%%%%%%%%%%%%%%%%%%%%%%%%%%
%%%%%%%%%%%%%%%%%%%%%%%%%%%%%%%%%%%%%%%%%%%%%%%%%%%%%%%%%%%%%%%%%%%%%%%%%%%%%%%%%%%%%
\section*{Acknowledgements}

H.S.P.M. is grateful to the Bundesministerium f\"ur Bildung und Forschung (BMBF) 
for financial support through project FKZ 50OF0901 (ICC HIFI \textit{Herschel}) 
during part of the present investigation. The measurements in K\"oln were supported 
by the Deutsche Forschungsgemeinschaft (DFG) through the collaborative research grants 
SFB~494 initially, and later SFB~956, project area B3. The portion of this work, 
which was carried out at the Jet Propulsion Laboratory, California Institute 
of Technology, was performed under contract with the National Aeronautics and 
Space Administration. The infrared spectra analyzed in the present study were 
recorded at the W.R. Wiley Environmental Molecular Sciences Laboratory, a national 
scientific user facility sponsored by the Department of Energy's Office of 
Biological and Environmental Research located at the Pacific Northwest National 
Laboratory (PNNL). PNNL is operated for the United States Department of Energy 
by the Battelle Memorial Institute under Contract DE-AC05-76RLO1830.

%%%%%%%%%%%%%%%%%%%%%%%%%%%%%%%%%%%%%%%%%%%%%%%%%%%%%%%%%%%%%%%%%%%%%%%%%%%%%%%%%%%%%
\appendix
%%%%%%%%%%%%%%%%%%%%%%%%%%%%%%%%%%%%%%%%%%%%%%%%%%%%%%%%%%%%%%%%%%%%%%%%%%%%%%%%%%%%%
\section*{Appendix A. Supplementary Material}

Supplementary data for this article are available on ScienceDirect (www.sciencedirect.com) 
and as part of the Ohio State University Molecular Spectroscopy Archives 
(http://library.osu.edu/sites/msa/jmsa\_hp.htm). 
The parameter, line, and fit files along with a readme file are provided as well as the 
full tables of the modeling of N$_2$- and self-broadening line widths of the $\nu _4$ 
band along with an explanation file. 
Supplementary data associated with this article can be found, in 
the online version, at http://dx.doi.org/10.1016/j.jms.2015.02.009. 

%% References


%%%  \bibliographystyle{elsarticle-num}
%%%  \bibliography{<your-bib-database>}



%%%%%%%%%%%%%%%%%%%%%%%%%%%%%%%%%%%%%%%%%%%%%%%%%%%%%%%%%%%%%%%%%%%%%%%%%%%%%%%%%%%%%
%%%%%%%%%%%%%%%%%%%%%%%%%%%%%%%%%%%%%%%%%%%%%%%%%%%%%%%%%%%%%%%%%%%%%%%%%%%%%%%%%%%%%
%%%%%%%%%%%%%%%%%%%%%%%%%%%%%%%%%%%%%%%%%%%%%%%%%%%%%%%%%%%%%%%%%%%%%%%%%%%%%%%%%%%%%

\end{document}